%% file: QCD_with_privacy_v15.tex
\documentclass[journal,10pt]{IEEEtran}

\input{preamble.tex}

\def\loadbibentry{}
\newcommand{\ARL}{\text{ARL}}
\newcommand{\WADD}{\text{WADD}}
\newcommand{\EWADD}{\text{EWADD}}

\DeclareRobustCommand{\stirling}{\genfrac\{\}{0pt}{}}

\newacronym{ARL}{ARL}{average run length}
\newacronym{WADD}{WADD}{worst-case average detection delay}
\newacronym{ADD}{ADD}{average detection delay}
\newacronym{PASC}{AOSC}{average observation-switching cost}
\newacronym{QCD}{QCD}{quickest change detection}
\newacronym{QCQP}{QCQP}{quadratically constrained quadratic program}
\title{Quickest Change Detection with Privacy Constraint}
\author{
Tze~Siong~Lau and Wee~Peng~Tay,~\IEEEmembership{Senior Member,~IEEE}
\thanks{This work was supported in part by the Singapore Ministry of Education Academic Research Fund Tier 2 grant MOE2018-T2-2-019 and by A*STAR under its RIE2020 Advanced Manufacturing and Engineering (AME) Industry Alignment Fund – Pre Positioning (IAF-PP) (Grant No. A19D6a0053). The computational work for this article was partially performed on resources of the National Supercomputing Centre, Singapore (\url{https://www.nscc.sg}).}%
\thanks{T.~S.~Lau and W.~P.~Tay are with the School of Electrical and Electronic Engineering, Nanyang Technological University, Singapore (e-mail: TLAU001@e.ntu.edu.sg, wptay@ntu.edu.sg).}
}
\begin{document}
\maketitle
\begin{abstract}
This paper considers Lorden's minimax quickest change detection (QCD) problem with a privacy constraint. The goal is to sanitize a signal to satisfy inference privacy requirements while being able to detect a change quickly. We show that the Generalized Likelihood Ratio (GLR) CuSum achieves asymptotic optimality with a properly designed sanitization channel. We formulate the design of this sanitization channel as an optimization problem, which is however challenging to solve. We propose relaxations to the optimization problem and develop algorithms to obtain a solution. We also consider the privacy-aware QCD problem under a decentralized framework and propose algorithms to solve the relaxed channel design problem under this framework.
\end{abstract}
%

\section{Introduction}
\label{sec:intro}


Quickest change detection (QCD) is the problem of sequentially detecting a change in the statistical properties of a signal. Given a sequence of independent and identically distributed (i.i.d.) observations $\{x_t:t\in\mathbb{N}\}$ with distribution $f$ up to an unknown change point $\nu$ and distribution $g\neq f$ after, the goal is to detect this change as quickly as possible, subject to false alarm constraints. Traditionally, applications of QCD can be found in manufacturing, in areas such as quality control\cite{lai1995sequential} where any change in the quality of products must be quickly detected. With the proliferation of low-cost sensors, QCD methods have also found applications in other areas such as fraud detection\cite{bolton02}, cognitive radio\cite{lai2008quickest} and power system line outage detection\cite{banerjee2014power}.

As sensor and computing technology become increasingly ubiquitous and powerful, it becomes easier for an adversary to infer sensitive information, such as lifestyle preferences and location information, from available data. In many practical QCD applications, rather than having one distribution, the distribution that generates the signal in the post-change regime belongs to a finite set $G=\{g_1,\ldots,g_{|G|}\}$. In some applications, the distribution of some sensitive information $U$ may depend on the post-change distribution. Thus, knowing which distribution $g\in G$ generates the signal in the post-change regime may reveal some information about $U$. One example of such an application is occupancy detection. In an Internet of Things (IoT) based occupancy detection system, the occupancy sensor continuously measures attributes such as infrared radiation, temperature, humidity and carbon dioxide levels, to quickly detect when a room becomes occupied so that certain functions like turning on the air conditioning system can be automated. A baseline distribution can be used to model the fluctuation of these attributes with time when the room is empty. However, fluctuations in these attributes when the room is occupied can reveal the number of people and the activity conducted in the room, leading to privacy leakage if an adversary has access to the raw attribute signals. The goal of privacy-aware QCD is to sanitize the attribute signals so that the change from a vacant to an occupied room can still be detected, while preserving the privacy of the occupants in the room. In practice, we are unable to know the adversary's intent and hence would like to sanitize the signal so that the largest improvement an adversary can achieve, over all possible queries, is controlled.

In other applications, $G$ may be partitioned into two sets, a private set $I_1$ and a public set $I_2$. We would like to sanitize the signal so that it is difficult for an adversary to deduce the distribution when a post-change distribution from the private set $I_1$ is generating the signal. On the other hand, we would like to retain the ability to deduce the distribution when a post-change distribution from the public set $I_2$ is generating the signal. One example of such an application is activity monitoring using wearables. In this application, we would like to quickly detect any change from a resting state to an active state. There are many possible active states such as walking, running, typing on a computer, and using a mobile phone. We would like to protect the privacy of some of these active states, like typing on a computer or using a mobile phone, while still being able to accurately track the other active states. Hence, the goal of privacy-aware QCD is to perform QCD while protecting the privacy of some active states and maintaining some distinguishability for the other active states.

In this paper, we address the QCD problem with multiple post-change distributions while maintaining a privacy constraint for two different privacy metrics. For each of the privacy metrics, we propose a signal sanitization algorithm and a stopping time that is able to identify the critical change quickly while preserving a pre-determined level of privacy.

\subsection{Related Work}

In the QCD problem with a single post-change distribution, when the pre- and post-change distributions are fully specified and the change point $\nu$ is unknown but deterministic, the Cumulative Sum (CuSum) test, developed by Page \cite{page54}, is optimal as the false alarm rate goes to zero. For the case where the post-change distribution is not fully specified, the GLR CuSum test is asymptotically optimal for the case of finite multiple post-change distributions. For a comprehensive overview of the QCD problem, we refer the reader to \cite{tartakovsky2014sequential,poor2009quickest,LauTayVee:J19,LauTay:J19} and the references therein.

In many applications, observations are obtained through measurements taken from several wireless sensors in the network and a fusion center decides if a change has taken place based on the information received from these sensors. Due to power and bandwidth constraints, the sensors are constrained to send messages belonging to a finite alphabet to the fusion center. This is an example in which the information for decision-making is \emph{decentralized}. The QCD problem with a decentralized framework was first introduced in \cite{veeravalli2001decentralized} and further studied in \cite{mei2005information,tartakovsky2008asymptotically,hadjiliadis2009one,banerjee2013decentralized} under various settings. All the aforementioned works on QCD do not consider any privacy constraints.

Existing work on protecting or quantifying privacy can be divided into two main categories: \emph{data} privacy and \emph{inference} privacy\cite{WangTIT2016,SunTay:C17,SunTay:J20b}. Data privacy refers to the protection of the sensors' raw information from being obtained by the fusion center. In contrast, inference privacy refers to the protection from an adversary's attempt to deduce properties of an underlying distribution. Privacy metrics proposed to quantify data privacy include local differential privacy \cite{Dwork2006,xiong2016randomized,duchi2013local}, $k$-anonymity\cite{GupRao:J17} and homomorphic encryption \cite{Boneh2005}. On the other hand, privacy metrics proposed to quantify inference privacy include average information leakage\cite{salamatian2013hide,khouzani2019generalized}, mutual information privacy \cite{WangTIT2016,song2017tutorial}, information privacy\cite{CalFaw:C12,SunTay:C16,SunTayHe:J18,HeTayHua:J19,SunTay:J20a}, maximal leakage privacy\cite{issa2016operational,issa2017operational}, local differential privacy, hypothesis testing adversary privacy\cite{LiOechtering:J19}, and compressive privacy\cite{KunSPM2017,SonWanTay:J20}. We refer the reader to \cite{SunTay:J20b,WangTIT2016} for a comprehensive discussion on the relationship between data and inference privacy. In this paper, we use maximal leakage privacy, and sequential hypothesis testing adversary privacy, which is the sequential analog to the privacy metric proposed in \cite{LiOechtering:J19}, to quantify the gain which the adversary obtains through the observation of the signal.

There are several works in the literature that present  privacy-preserving frameworks for different signal processing tasks\cite{SunTay:J20a,rassouli2020,liao2017hypothesis,cummings2018differentially}. In \cite{cummings2018differentially}, the authors developed differentially private algorithms, which assume that the adversary knows all entries of the database except one, for the purpose of change-point detection. Unlike \cite{cummings2018differentially}, we consider a weaker form of privacy as this assumption may be too strong for some applications. Furthermore, we provide theoretical guarantees on the average run length to false alarm and the worst-case detection delay, which are more relevant to the QCD task as compared to guarantees on the accuracy of the estimated change-point provided by \cite{cummings2018differentially}, which are more suited for the change-point detection task. In \cite{rassouli2020}, the sanitization channel is designed for general signal processing tasks while preserving privacy, where the utility of the sanitized signal is measured using general information theoretic quantities such as mutual information, minimum mean-square error (MMSE), and probability of error.  In \cite{liao2017hypothesis,SunTay:J20a}, the authors consider a fixed sample size problem of hypothesis testing while preserving privacy, where the utility of the sanitized signal is measured by the Bayes error or the Type II error of the test. 

\subsection{Our Contributions}

In this paper, we consider the problem of optimizing QCD performance while preserving privacy. Furthermore, unlike the papers mentioned above, we consider sequential signals. It is thus possible for the adversary to obtain an arbitrarily large number of samples to improve his guess. Our main contributions are summarized as follows:
\begin{itemize}
\item We formulate the QCD problem with two privacy constraints, the maximal leakage privacy metric and the sequential hypothesis testing privacy metric.
\item We show that the
GLR CuSum stopping time together with a properly designed sanitization
channel is asymptotically optimal.
\item We propose relaxations and algorithms for both the centralized and decentralized versions of the QCD problem with privacy constraints.
\end{itemize}
A preliminary version of this paper was presented in \cite{lau2020quickest}.	

The rest of this paper is organized as follows. In \cref{sec:problem}, we present our signal model and problem formulation. We derive the asymptotic optimality of the GLR CuSum stopping time and formulate an optimization problem to design the optimal sanitization channel in \cref{sec:asym_opt}. We propose relaxations to the channel design problem in \cref{sec:relaxation} and present methods to solve the relaxed channel design problem in \cref{sec:centralized_algorithms}. We present the signal and sanitization model for the decentralized privacy-aware QCD problem in \cref{sec:iid} and propose methods to solve the corresponding relaxed channel design problem in \cref{sec:decentralized_algorithms}. Results from numerical experiments are presented in \cref{sec:numerical}. We conclude in \cref{sec:conclusion}.

\section{Problem formulation}\label{sec:problem}
Let $\calX$ be a measurable space, where $\calX$ is a finite alphabet. We consider a sequence of random variables $X_1, X_2, \ldots$ taking values in $\calX$ and \gls{iid} according to different distributions before and after an unknown change point. Let $f$ be the pre-change distribution and $G=\{g_1,g_2,\ldots,g_{|G|}\}$ be the set of possible post-change distributions on $\calX$ such that $f\neq g_i$ for all $i\in\{1,2,\ldots,|G|\}$. Let $I$ be a random variable on the indices of $G$ with distribution $p_I$.We assume that the sequence of random variables $X_1,X_2,\ldots$ satisfy the following:
\begin{align}\label{eqn:signalmodel}
\begin{cases}
X_{t} \sim f \quad \text{i.i.d.\ for all $t< \nu$},\\
X_{t} \sim g_{i} \quad \text{i.i.d.\ for all $t\geq \nu$},\\
\end{cases}
\end{align}
where $\nu\geq 0$ is an unknown but deterministic change point, $i$ is the realization of the random variable $I$ which remains fixed for all $t\geq\nu$. We further assume that an adversary is interested in obtaining information about a random variable $U$, unknown to the data curator, which takes on values in a finite set $\calU$, and that $U$ can be expressed as a randomized function of $I$, i.e., the identity of the post-change distribution informs us about $U$.

We restrict our analysis to memoryless privacy mechanisms. A privacy mapping or sanitization channel $q$ maps an observation $X \in \calX$ to a random variable $Y \in \calY$, where $\calY$ is a discrete alphabet such that $|\calY| \leq |\calX|$. The sanitization channel $q$ can be represented by the conditional probability $\P(Y=y \ | X=x)$. Let $T_q$ be a column-stochastic matrix with $[T_q]_{y,x}=\P(Y=y){X=x}$ where $ [T]_{y,x}$ denotes the $(y,x)$ entry of a matrix $T$. Likewise, we represent a distribution $h$ on $\calX$ as a column vector with $[h]_x=h(x)$ and similarly for a distribution on $\calY$.

At each time $t$, we apply a sanitization channel $q$ to obtain $Y_t=q(X_t)$. The sanitized signal $Y_t$ is generated i.i.d.\ by the distribution $\widetilde{f}=T_{q}f$ in the pre-change regime and by the distribution $T_{q}g_i$ in the post-change regime, for some $1\leq i\leq |G|$. For a fixed $q$, we let 
\begin{align*}
\widetilde{G}=\{T_{q}g_i\ :\ \text{$1\leq i\leq|G|$}\}
\end{align*} 
to be the set of possible post-change distributions. Since the distributions $T_{q}g_i$ may not be distinct, we have $|\widetilde{G}|\leq |G|$.

In this paper, we study the QCD problem with privacy constraints using two different privacy metrics. We assume that, at each time $t$, the adversary knows the pre-change distribution $f$, the set $G$ of post-change distributions, change-point $\nu$, the sanitization channel $q$, the current and all previous observations $Y^{1:t}=\{Y_1,Y_2,\ldots,Y_t\}$.

The first privacy metric we consider is maximal leakage, first proposed in \cite{issa2016operational}, to quantify the amount of information leakage an adversary is able to gain from observing the signal $\{Y_t\ :\ t\in\mathbb{N}\}$, where $\bbN$ is the set of positive integers. Given two random variables $A \in\calA$ and $B\in\calB$, the maximal leakage from $A$ to $B$ is defined as
\begin{align}\label{eqn:maxleak}
\calL_{\text{max}}(A\to B)=\sup_{U-A-B-\widehat{U}}\log\frac{\P(\widehat{U}=U)}{\max_{u\in\calU}\P(U=u)},
\end{align}
where $U - A - B - \hat{U}$ denotes a Markov chain and the supremum is taken over all such Markov chains. The quantity $\calL_{\text{max}}(A\to B)$ can be interpreted as the maximum gain in bits (if $\log$ is base 2) an adversary can achieve in guessing $U$ by observing $B$, where $U$ is a randomized function of $A$. The expression in \eqref{eqn:maxleak} is equivalent\cite{issa2016operational} to 
\begin{align}\label{eqn:maximal_leakage_closed_form}
\calL_{\text{max}}(A\to B)=\log\left(\sum_{a\in \calA}\max_{b\in\calB}\P(B=b){A=a}\right).
\end{align}
Thus, at each time $t$, the maximum gain in bits an adversary can achieve in guessing $U$ is $\calL_{\text{max}}(I\to Y^{\nu:t})$. 

The second privacy metric we consider is the \emph{sequential hypothesis testing privacy} metric. It quantifies the amount of gain an adversary is able to achieve by performing a sequential hypothesis test. Using this privacy metric, we are able to protect a subset of the post-change hypotheses from the inference of an adversary while ensuring the distinguishability of the rest of the post-change hypotheses. Given a sanitization channel $q$ and a partition $I_1\cup I_2$ of the index set of $G$ with $|I_1|>1$, such that $I_1$ is the set of indices of the post-change hypotheses to be protected, we define 
\begin{align*}
\calK_1(T_q)=\max_{i\in I_1} \min_{j\in I_1}\KLD{	T_qg_{i}}{T_qg_{j}},\\
\calK_2(T_q)=\min_{i\in I_2} \min_{j\in I_1\cup I_2}\KLD{T_qg_{i}}{T_qg_{j}},
\end{align*}
where $\KLD{\cdot}{\cdot}$ is the Kullback-Leibler (KL) divergence.

Using standard results from sequential hypothesis testing\cite[Theorem 4.3.1]{tartakovsky2014sequential}, assuming that the adversary is willing to accept a misclassification rate of $\eta$, the expected number of samples required to identify a distribution with index in $I_1$ is \emph{at least} $|\log \eta|/\calK_1(T_q)$ asymptotically  as $\eta\to 0$. Similarly, the expected number of samples required to identify a distribution with index in $I_2$ is \emph{at most} $|\log \eta|/\calK_2(T_q)$ asymptotically as $\eta\to 0$.

For a fixed sanitization channel $q$, the QCD problem is to detect a change in distribution as quickly as possible by observing the sanitized signal $Y_1=q(X_1),Y_2=q(X_2),\ldots$, while keeping the false alarm rate low. In a typical sequential change detection procedure, at each time $t$, a test statistic $S(t)$ is computed based on the observations $Y_1,\ldots,Y_t$ up to time $t$, and the observer decides that a change has occurred at a stopping time $\tau=\inf\{t:S(t)>b\}$, which is the first $t$ such that $S(t)$ exceeds a pre-determined threshold $b$. The QCD performance of a stopping time $\tau$ can be quantified using the trade-off between two quantities, the average run length to false alarm, $\ARL(\tau)$, and the expected worst-case average detection delay, $\EWADD(\tau)$, defined as
\begin{align*}
\ARL(\tau)&=\E{\infty}[\tau],\\
\EWADD(\tau)&=\E[\WADD_I(\tau)], \\
\WADD_i(\tau)&=\sup_{\nu\geq 1}\esssup \E{\nu,i}[(\tau-\nu_c+1)^+|Y_1^{\nu_c-1}],
\end{align*}
where $\esssup$ is the essential supremum operator, $\E_{\nu,i}$ is the expectation operator assuming the change-point is at $\nu$ with post-change distribution $g_i$, and $\E_{\infty}$ is the expectation operator assuming the change does not occur. We should note that typically, there are several sanitization channels $q$ that satisfy a privacy constraint. As the pre and post-change distributions, $\{\widetilde{f}\}\cup\widetilde{G}$, vary with the sanitization channel $q$, we expect that the QCD performance varies with $q$ as well. It is then important for us to select the sanitization channel $q$ that provides the best QCD performance while satisfying the privacy constraint.

Our privacy-aware QCD problem can be formulated as an optimization problem as follows: given a privacy admissible set $\calQ$ and an average run length requirement $\gamma$, we seek a stopping time $\tau$, and a sanitization channel $q$ such that they are optimal solutions to the following problem:
\begin{equation}\label{eqn:optimize_formulation}
\begin{aligned}
& \underset{\tau,q}{\text{minimize}}
& & \EWADD(\tau) \\
& \text{subject to}
& & \ARL(\tau)\geq\gamma,\\
& &&q\in \calQ
\end{aligned}
\end{equation}
where under the first privacy metric, the privacy admissible set $\calQ$ is defined as 
\begin{align*}
\calQ=\{q\ :\ \sup_{0\leq\nu\leq t}\calL_{\text{max}}(I\to Y^{\nu:t})\leq\epsilon\}
\end{align*}
for some given privacy budget $\epsilon>0$, and under the second privacy metric,
\begin{align*}
\calQ=\{q\ :\ \calK_1(T_q)\leq \epsilon_1,\ \calK_2(T_q)\geq \epsilon_2\}
\end{align*}
for some given privacy budget $\epsilon_1$ and distinguishability level $\epsilon_2>0$.

\section{Asymptotic optimality}\label{sec:asym_opt}
In this section, we present the GLR CuSum stopping time for the privacy-aware QCD problem and study its asymptotic properties as $\gamma\to\infty$. First, we note that the minimization over the sanitization channel $q$ and stopping time $\tau$ can be decoupled in Problem~\cref{eqn:optimize_formulation}. For a fixed $q\in\calQ$, we define the GLR CuSum stopping time $\omega_{q}$ and the GLR CuSum test statistic $S(t)$ as follows
\begin{align*}
&\ \  \ \omega_{q}=\inf\left\{t\ :\ S(t)\geq b\right\},\\
&\ \ \ S(t)=\max_{1\leq j\leq |\widetilde{G}|}S_j(t),\\
&\begin{cases}
S_j(t)&=\max\left(S_j(t-1)+\log\frac{\widetilde{g}_j(y_t)}{\widetilde{f}(y_t)},0\right)\\
S_j(0)&=0,
\end{cases}\text{for $1\leq j\leq |\widetilde{G}|$. }
\end{align*}
When the signal $\{\bX_t\ :\ t\in\mathbb{N}\}$ is sanitized using the channel $q$,
the GLR CuSum stopping time $\omega_{q}$ is asymptotically optimal\cite{lorden71} for the following problem: 
\begin{equation}\label{eqn:optimize_formulation_lorden}
\begin{aligned}
& \underset{\tau}{\text{minimize}}
& & \EWADD(\tau)\\
& \text{subject to} &&\ARL(\tau)\geq\gamma,
\end{aligned}
\end{equation}
with the asymptotic $\ARL$-$\EWADD$ trade-off given as
\begin{align*}
\EWADD(\omega_{q})=\E[\frac{\log \gamma}{\KLD{T_{q}g_I}{T_{q}f}}](1+o(1))
\end{align*}
as $\gamma\to\infty$, where the expectation is taken with respect to $I$.
Let $q^*$ be an optimal solution to the following problem:
\begin{equation}\label{eqn:optimize_formulation_asymptotic}
\begin{aligned}
& \underset{q}{\text{minimize}}
& & \E[\frac{1}{\KLD{T_{q}g_I}{T_{q}f}}], \\
&\text{subject to} && q\in\calQ.
\end{aligned}
\end{equation}
Using similar arguments from \cite{tartakovsky2014sequential,lau2017optimal}, it can be shown that $\omega_{q^*}$ is asymptotically optimal for Problem~\cref{eqn:optimize_formulation} as $\gamma\to \infty$.

We call Problem~\cref{eqn:optimize_formulation_asymptotic} the channel design problem and note that it is challenging to solve for several reasons. First, the objective function is neither concave nor convex. Second, it is difficult to obtain a closed form expression for the  maximal leakage privacy constraint since the alphabet size of $Y^{\nu:t}$ increases quickly as $t\to\infty$. The sequential hypothesis testing privacy constraints are also neither concave nor convex. In the next section, we present relaxations of the constraint and objective function of Problem~\cref{eqn:optimize_formulation_asymptotic} to improve its computational tractability.

\section{Relaxation of the Channel Design Problem}\label{sec:relaxation}
\subsection{Relaxation of the Objective Function}
In this subsection, we provide a relaxation of the objective function in Problem~\cref{eqn:optimize_formulation_asymptotic}. We propose to relax the objective function using Jensen's inequality:
\begin{align*}
\E[\frac{1}{\KLD{T_{q}g_I}{T_{q}f}}] \geq \frac{1}{\E[\KLD{T_{q}g_I}{T_{q}f}]} .
\end{align*}
By replacing the objective function with its lower bound, we obtain the following relaxed problem: 
\begin{equation}\label{eqn:optimize_formulation_asymptotic_relaxed_objective}
\begin{aligned}
& \underset{q}{\text{maximize}}
& & \E[\KLD{T_{q}g_I}{T_{q}f}] \\
& \text{subject to} &&q\in\calQ.
\end{aligned}
\end{equation}
For a fixed post-change distribution $g_i$, with $1\leq i\leq N$, the expected rate of growth of $S(t)$ in the post-change regime is given as $\KLD{T_q g_i}{T_qf}$. Thus, we can interpret the new objective function as the expected rate of growth of $S(t)$ averaged over the different post-change distributions. Since the stopping time $\omega_{q}$ declares that a change has taken place when the test statistic $S(t)$ exceeds a pre-defined threshold $b$, this means that, heuristically, a larger expected rate of growth of $S(t)$ gives a smaller $\EWADD$. This intuition agrees with the relaxed problem \cref{eqn:optimize_formulation_asymptotic_relaxed_objective} obtained by replacing the objective function with its lower bound ${1}/{\E[\KLD{T_{q}g_I}{T_{q}f}]}$ in Problem~\cref{eqn:optimize_formulation_asymptotic}.
\subsection{Relaxation of Privacy Constraints}
\subsubsection{Maximal Leakage privacy}\label{subsec:ml_constraint} 
In this subsection, we focus on the relaxation of the constraint $q\in\calQ$ when the privacy metric is the maximal leakage privacy. Under the maximal leakage privacy metric, Problem~\cref{eqn:optimize_formulation_asymptotic_relaxed_objective} becomes:
\begin{equation}\label{eqn:optimize_formulation_asymptotic_relaxed_objective_ml}
\begin{aligned}
& \underset{q}{\text{maximize}}
& & \E[\KLD{T_{q}g_I}{T_{q}f}] \\
& \text{subject to} &&\sup_{0\leq\nu\leq t}\calL_{\text{max}}(I\to Y^{\nu:t})\leq\epsilon .
\end{aligned}
\end{equation}
As it is difficult to obtain a closed form expression for the maximal leakage $\calL_{\max}(I\to Y^{\nu:t})$, we approximate it using an upper bound which is easily computable. Let $J$ be a random variable on the indices of $\widetilde{G}=\{\widetilde{g}_1,\widetilde{g}_2,\ldots,\widetilde{g}_{|\widetilde{G}|}\}$, such that
\begin{align*}
\P(J=j){I=i}&=\begin{cases}
1 \quad\text{if $\widetilde{g}_j= T_{q}g_i$,}\\
0\quad\text{otherwise.}
\end{cases}
\end{align*}
Let $U$ be a randomized function of $I$. According to our signal model, we have the following factorization,
\begin{align*}
P_{J,I,Y^{\nu:t},U}&=P_{J,I,U}\ P_{Y^{\nu:t}|I,J,U}\\
&=P_{I,J,U}\ P_{Y^{\nu:t}| J}\\
&=P_I\ P_{J | I}\ P_{U |I}\ P_{Y^{\nu:t} | J},
\end{align*}
where we use $P_X$ to denote the probability mass function (pmf) of $X$ and  $P_{X|Y}$ to denote the conditional pmf of $X$ given $Y$.
The following proposition provides the motivation to relax the privacy constraint in Problem~\eqref{eqn:optimize_formulation_asymptotic_relaxed_objective_ml} to $\calL_{\text{max}}(I\to J)\leq \epsilon$.
\begin{Proposition}\label{prop:upperbound_max_leakage}
For any $t\in\mathbb{N}$, we have 
$\calL_{\text{max}}(I\to Y^{\nu:t})\leq\calL_{\text{max}}(I\to J).$ Hence, we have
\begin{align*}
\sup_{0\leq \nu\leq t}\calL_{\text{max}}(I\to Y^{\nu:t})\leq\calL_{\text{max}}(I\to J).
\end{align*}
\end{Proposition}
\begin{IEEEproof}
See \cref{sec:AppProp1}.
\end{IEEEproof}

Replacing the privacy constraint $	\sup_{0\leq \nu\leq t}\calL_{\text{max}}(I\to Y^{\nu:t})\leq \epsilon$ in \cref{eqn:optimize_formulation_asymptotic_relaxed_objective_ml} with $\calL_{\text{max}}(I\to J)\leq \epsilon$, we obtain the relaxed channel design problem:
\begin{equation}\label{eqn:optimize_relaxed_formulation}
\begin{aligned}
& \underset{q}{\text{maximize}}
& & \E[\KLD{T_{q}g_I}{T_{q}f}]\\
& \text{subject to}
& & \calL_{\text{max}}(I\to J)\leq\epsilon.
\end{aligned}
\end{equation}
\cref{prop:upperbound_max_leakage} guarantees that any solution to Problem~\cref{eqn:optimize_relaxed_formulation} satisfies the original privacy constraint $\sup_{0\leq \nu\leq t}\calL_{\text{max}}(I\to Y^{\nu:t})\leq\epsilon$.

\subsubsection{Relaxation of the sequential hypothesis testing privacy constraint}\label{subsec:sht_constraint} 
In this subsection, we focus on the relaxation of the constraint $q\in\calQ$ when the privacy metric is the sequential hypothesis testing privacy metric. Under the sequential hypothesis testing privacy metric, Problem~\cref{eqn:optimize_formulation_asymptotic_relaxed_objective} becomes:
\begin{equation}\label{eqn:optimize_formulation_asymptotic_relaxed_objective_sht}
\begin{aligned}
& \underset{q}{\text{maximize}}
& & \E[\KLD{T_{q}g_I}{T_{q}f}] \\
& \text{subject to} &&\calK_1(T_q)\leq \epsilon_1,\ \calK_2(T_q)\geq \epsilon_2.
\end{aligned}
\end{equation}
Problem~\cref{eqn:optimize_formulation_asymptotic_relaxed_objective_sht} is non-convex as the constraints $\calK_1(T_q)\leq \epsilon_1$ and $\calK_2(T_q)\geq \epsilon_2$ are non-convex. This makes it difficult to have any theoretical guarantees of the global optimality of solutions found for Problem~\cref{eqn:optimize_formulation_asymptotic_relaxed_objective_sht}. We focus on a restricted sanitization model in order to improve the computational tractability of Problem~\cref{eqn:optimize_formulation_asymptotic_relaxed_objective_sht}. Let $\calC=\{q_1,q_2,\ldots,q_n\}$ be a finite set of sanitization channels with column-stochastic matrices $T_1,\ldots,T_n$. At each time instance $t$, we assume that the observer obtains an observation $Y_t=(Z_t,A_t)$ where $Z_t=q_{A_t}(X_t)$ is a randomized function of the random variable $X_t$ under the sanitization channel $q_{A_t}$. We further assume that $\{A_t\}_{t\in\mathbb{N}}$ are i.i.d. generated with distribution $\phi$ on $\{1,2,\ldots,n\}$. Under the restricted sanitization model, rather than designing the sanitization channel $q$, we design the distribution $\phi$ that samples a sanitization channel $q_{A_t}$ from  $\calC$ at each time instance $t$ and use $q_{A_t}$ to sanitize the signal.

For a fixed set of sanitization channels $\calC=\{q_1,q_2,\ldots,q_n\}$  and a fixed distribution $\phi$, the asymptotic $\ARL$-$\EWADD$ trade-off of the GLR CuSum stopping time under the restricted sanitization model can be derived, using similar arguments from \cite{lau2017optimal}, to be
\begin{align*}
\EWADD(\omega_{q})=\E[\frac{\log \gamma}{\sum_{c=1}^n\phi(c)\KLD{T_c g_I}{T_c f}}](1+o(1)),
\end{align*}
as $\gamma\to\infty$. Furthermore, we have
\begin{align*}
\E[\KLD{T_{q}g_I}{T_{q}f}]=\sum_{c=1}^n \phi(c)\E[\KLD{T_{c}g_I}{T_{c}f}].
\end{align*} Thus, Problem~\cref{eqn:optimize_formulation_asymptotic_relaxed_objective_sht} becomes
\begin{equation}\label{eqn:optimize_formulation_asymptotic_relaxed_objective_sht_linearized}
\begin{aligned}
& \underset{\phi}{\text{maximize}}
& & \sum_{c=1}^n \phi(c)\E[\KLD{T_{c}g_I}{T_{c}f}] \\
& \text{subject to} &&\max_{i\in I_1} \min_{j\in I_1}\sum_{c=1}^n\phi(c)\KLD{T_cg_i}{T_cg_j}\leq \epsilon_1,\\
& & &\sum_{c=1}^n \phi(c)=1,\\
& & &\phi(c)\geq 0, \quad\text{$c\in\{1,\ldots,n\}$,}\\
& & &\sum_{c=1}^n\phi(c)\KLD{T_cg_i}{T_cg_j}\geq \epsilon_2\\
& & &\text{for $i\in I_2$ and $j\in I_1\cup I_2$}.
\end{aligned}
\end{equation}
Since $\KLD{T_{c}g_I}{T_{c}f},\KLD{T_{c}g_i}{T_{c}g_j}$ can be pre-computed for all $c\in\{1,\ldots,n\}$ and $i,j\in \{1,\ldots,|G|\}$, Problem~\cref{eqn:optimize_formulation_asymptotic_relaxed_objective_sht_linearized}  without the constraint 
\begin{align*}
\max_{i\in I_1} \min_{j\in I_1}\sum_{c=1}^n\phi(c)\KLD{T_cg_i}{T_cg_j}\leq \epsilon_1
\end{align*} is a linear program. In the next section, we show that Problem~\cref{eqn:optimize_formulation_asymptotic_relaxed_objective_sht_linearized} can be expressed as a mixed-integer linear program (MILP).

\section{Algorithms for Privacy-Aware QCD}\label{sec:centralized_algorithms}
In this section, methods that provide globally and locally optimal solutions for Problems~\cref{eqn:optimize_relaxed_formulation,eqn:optimize_formulation_asymptotic_relaxed_objective_sht_linearized} are presented.
\subsection{Maximal Leakage Privacy}\label{subsec:algorithms_ml}
\subsubsection{Exact Method}
From \eqref{eqn:maximal_leakage_closed_form}, $2^{\calL_{\text{max}}(I\to J)}$ is an integer since the conditional probability $\P(J=j){I=i}\in\{0,1\}$ is an integer. Hence, the constraint $\calL_{\text{max}}(I\to J)\leq \epsilon$ in Problem~\cref{eqn:optimize_relaxed_formulation} is equivalent to
\begin{align}\label{eqn:L(IJ)=m}
\sum_j\max_i\P(J=j){I=i}\leq m,
\end{align}
where $m=\lfloor2^\epsilon\rfloor$. There are at most $\stirling{|G|}{m}\leq m^{|G|}$ different conditional pmfs $P_{J|I}$ satisfying \eqref{eqn:L(IJ)=m}, where the Stirling number of the second kind $\stirling{a}{b}$ counts the number of ways to partition a set of $a$ labeled objects into $b$ nonempty unlabeled subsets\cite{riordan2012introduction}. For each conditional pmf $P_{J|I}$, we solve the following problem:
\begin{equation}\label{eqn:N=1,epsilon>0,cases}
\begin{aligned}
& \underset{q}{\text{maximize}}
& & \E[\KLD{T_{q}g_i}{T_{q}f}] \\
& \text{subject to}
& &  T_{q}g_i=\widetilde{g}_j \\
& & &\text{for all $i,j$ such that $\P(J=j){I=i}=1$}.\\
\end{aligned}
\end{equation}
For a fixed conditional pmf $P_{J|I}$, Problem~\eqref{eqn:N=1,epsilon>0,cases} is maximizing a convex function over a convex bounded polytope. Therefore, an extreme point achieves the maximum value, and we are able to solve Problem~\eqref{eqn:N=1,epsilon>0,cases} by enumerating over the finite number of extreme points of the convex bounded polytope defined by the linear constraints of \eqref{eqn:N=1,epsilon>0,cases}. A globally optimal solution for Problem~\cref{eqn:optimize_relaxed_formulation} can be obtained by enumerating over all conditional pmfs $P_{J|I}$ represented by a zero-one matrix satisfying \eqref{eqn:L(IJ)=m}, and solving Problem~\eqref{eqn:N=1,epsilon>0,cases} for each of these conditional pmfs.

However, we still need to solve at least exponentially many convex maximization problems with respect to the number of post-change distributions $|G|$, since $\stirling{|G|}{m}\sim\frac{m^{|G|}}{m!}$ as $|G|\to \infty$. This may be computationally undesirable when $|G|$ is large.

\subsubsection{Augmented Lagrangian Method}
We further relax the channel design problem by relaxing the discrete constraint \eqref{eqn:L(IJ)=m}. This relaxation allows the application of the augmented Lagrangian method for cases where the exact method is computationally undesirable.
The constraint \eqref{eqn:L(IJ)=m} is equivalent to 
$
|\{T_qg_i \ :\ 1\leq i\leq |G|\}|\leq m.
$
In order to count the number of distinct elements in the set $\{T_qg_i \ :\ 1\leq i\leq |G|\}$, we can use the following
continuous approximation,\begin{align*}
&|\{T_qg_i \ :\ 1\leq i\leq |G|\}|\\
&=1+\sum_{i=2}^{|G|}\prod_{j=1}^{i-1} \mathbf{1}_{T_qg_i\neq T_qg_j}\\
&\approx 1+\sum_{i=2}^{|G|}\prod_{j=1}^{i-1} \left(\frac{1}{2}+ \frac{1}{\pi}\arctan (k\|T_qg_i-T_qg_j\|_1)\right),
\end{align*}
where $k$ is a chosen to be large and $\|\cdot\|_1$ refers to the $L_1$ norm. Putting this back into Problem~\cref{eqn:optimize_relaxed_formulation}, we obtain the following continuous optimization problem:
\begin{equation}\label{eqn:optimize_continuous_relaxed_formulation}
\begin{aligned}
&\underset{q}{\text{maximize}}\ \E[\KLD{T_{q}g_n}{T_{q}f}] \\
& \text{subject to} \\
& \sum_{i=2}^{|G|}\prod_{j=1}^{i-1} \left(\frac{1}{2}+ \frac{1}{\pi}\arctan (k\|T_qg_i-T_qg_j\|_1)\right)\leq m-1, 
\end{aligned}
\end{equation}
for which an augmented Lagrangian Solver\cite{pyopt-paper,conn2013lancelot} can be used to obtain locally optimal solutions\cite{fernandez2012local}.

\subsection{Sequential Hypothesis Testing Privacy}\label{subsec:algorithms_sht}
In this subsection, we show that Problem~\cref{eqn:optimize_formulation_asymptotic_relaxed_objective_sht_linearized} is equivalent to a MILP. First, we require the following proposition.
\begin{Proposition}\label{prop:equiv_conditions}
For  distribution $\phi$ on $\{1,\ldots,n\}$, $\phi$ satisfies 
\begin{align}\label{eqn:equiv_prop_1}
\max_{i\in I_1} \min_{j\in I_1}\sum_{c=1}^n\phi(c)\KLD{T_cg_i}{T_cg_j}\leq \epsilon_1
\end{align} 
if and only if there exist functions
\begin{align*}
\xi&:I_1\to \mathbb{R},\\
\delta&:I_1\times I_1 \to\{0,1\}
\end{align*}
such that
\begin{align}
&\xi(i)\leq \epsilon_1,\label{eqn:eqn:equiv_prop_2_1}\\
&\sum_{j\in I_1}\delta(j,i)=1,\label{eqn:eqn:equiv_prop_2_2}\\
&\sum_{c=1}^n\phi(c)\KLD{T_cg_i}{T_cg_j}\geq \xi(i),\label{eqn:eqn:equiv_prop_2_3}\\
&\sum_{c=1}^n\phi(c)\KLD{T_cg_a}{T_cg_j}\leq \xi(i)+(1-\delta(j,i))M,\label{eqn:eqn:equiv_prop_2_4}
\end{align}
where 
$M=\max_{c\in\{1,\ldots,n\}}\max_{i,j\in I_1}\KLD{T_cg_i}{T_cg_j},$ and $i,j\in I_1$.
\end{Proposition} 
\begin{IEEEproof}
See \cref{sec:AppProp2}.
\end{IEEEproof}

By \cref{prop:equiv_conditions}, Problem~\cref{eqn:optimize_formulation_asymptotic_relaxed_objective_sht_linearized} is equivalent to the following MILP:
\begin{align}\label{eqn:optimize_formulation_asymptotic_relaxed_objective_sht_MILP}
&\underset{\phi}{\text{maximize}} \ \sum_{c=1}^n \phi(c)\E[\KLD{T_{c}g_I}{T_{c}f}] \nn
&\text{subject to}  \nn
& \sum_{c=1}^n\phi(c)\KLD{T_cg_i}{T_cg_j}\geq \epsilon_2 \nn
& \quad\quad\text{for $i\in I_2$ and $j\in\{1,\ldots,|G|\}$},\nn
& \xi(i)\leq \epsilon_1\quad\text{ for $i\in I_1$},\nn
& \delta(j,i)\in\{0,1\}\text{ for $i,j\in I_1$},\\
& \sum_{c=1}^n \phi(c)=1\nn
& \phi(c)\geq 0 \quad\text{$c\in\{1,\ldots,n\}$,}\nn
& \sum_{j\in I_1}\delta(j,i)=1\text{ for $i\in I_1$},\nn
& \sum_{c=1}^n\phi(c)\KLD{T_cg_i}{T_cg_j}\geq \xi(i) \ \text{for $i,j\in I_1$},\nn
& \sum_{c=1}^n\phi(c)\KLD{T_cg_i}{T_cg_j}\nn
& \quad\quad\leq \xi(i)+(1-\delta(j,i))M\quad\text{for $i,j\in I_1$}.\nonumber
\end{align}
A global optimal solution to Problem~\cref{eqn:optimize_formulation_asymptotic_relaxed_objective_sht_MILP} can be obtain using branch-and-bound methods on a linear program solver\cite{cvx}.

\section{Decentralized QCD with Independent Sensor observations}\label{sec:iid}
We assume that the sequence of random variables $X_1,X_2,\ldots$ satisfy the observations obtained at each sensor at any time instance are independent before the change point, and conditionally independent given $I$ after the change point. The observation obtained by the $k$-th sensor at time $t$, $X_{k,t}$,  taking values in $\calX$, satisfy the following:
\begin{align}\label{eqn:iidsignalmodel}
\begin{cases}
X_{k,t} \sim f_k \quad \text{i.i.d. for all $t< \nu$},\\
X_{k,t} \sim g_{k,i} \quad \text{i.i.d. for all $t\geq \nu$},\\
\end{cases}
\end{align}	
 for $k\in\{1,\ldots,K\}$. The observations $X_{1,t},\ldots,X_{K,t}$ are mutually independent, $\nu\geq 0$ is an unknown but deterministic change point and $i$ is the realization of the random variable $I$ which remains fixed for all $t\geq\nu$.  We denote the marginal distribution of $X_{k,t}$ under $f$ and $g_i$ as $f_k$ and $g_{k,i}$ respectively. 

For the task of privacy-aware decentralized QCD, we apply a memoryless privacy mechanism locally at each sensor $k$ for $k\in\{1,\ldots,K\}$. For each sensor $k$, a local privacy mapping or sanitization channel $q_k$ maps the observation $X \in \calX$ obtained at sensor $k$ to a random variable $Y \in \calY$, where $\calY$ is a discrete alphabet such that $|\calY| \leq |\calX|$. The local privacy mechanism $q_k$ can be represented by a conditional probability $\P(Y=y \ | X=x)$. Let $T_{q_k}$ be a column-stochastic matrix with $[T_{q_k}]_{y,x}=\P(Y=y){X=x}$ where $ [T_{q_k}]_{y,x}$ denotes the $(y,x)$ entry of the matrix $T_{q_k}$. Likewise, we represent a distribution $h$ on $\calX$ as a column vector with $[h]_{x}=h(x)$ and similarly for a distribution on $\calY$.

At each time $t$ and sensor $k$, we apply the local sanitization channel $q_k$ to obtain $Y_{k,t}=q_k(X_{k,t})$. The sanitized signal $Y_{k,t}$ is generated i.i.d.\ by the distribution $T_{q_k} f_k$ in the pre-change regime and by the distribution $T_{q_k} g_{k,i}$ in the post-change regime, for some $1\leq i\leq |G|$. For a fixed set of sanitization channels $\{q_1,\ldots,q_K\}$, we let 
$\widetilde{G}_{\{q_1,\ldots,q_K\}}=\{\widetilde{g}_1,\ldots,\widetilde{g}_{|G|}\}$ 
to be the set of possible post-change distributions where $\widetilde{g}_j$ is the  post-change distribution generating the sanitized signal by applying the set of sanitization channels $\{q_1,\ldots, q_K\}$ to the observations generated by $g_i$ for some $i\in\{1,\ldots,|G|\}$.

\subsection{Algorithms for Decentralized Privacy-Aware QCD}\label{sec:decentralized_algorithms}
For the task of decentralized privacy-aware QCD , the sanitization channels are only allowed to use local observations to achieve sanitization of the signal. This introduces additional constraints on the structure of the sanitization channel $q$. In this section, we present algorithms for solving Problems~\cref{eqn:optimize_relaxed_formulation,eqn:optimize_formulation_asymptotic_relaxed_objective_sht_linearized} for the task of decentralized privacy-aware QCD.
\subsection{Maximal Leakage Privacy}\label{subsec:decen_algorithms_ml}
In this subsection, we present the Local Exact Method which solves Problem~\cref{eqn:optimize_relaxed_formulation} exactly and has computational complexity that scales linearly with respect to the number of sensors $K$. 

First, for each conditional pmf $P_{J|I}$ satisfying 
$
\P(J=j){I=i}\in\{0,1\}
$ for all $1\leq i\leq |G|$, $1\leq j \leq |\widetilde{G}|$ and $\calL(I\to J)\leq \epsilon$, we solve the following problem:
\begin{equation}\label{eqn:cases}
\begin{aligned}
& \underset{q_k}{\text{maximize}}
& & \E[\KLD{T_{q_k} g_{k,I}}{T_{q_k} f_k}] \\
& \text{subject to}
& &  T_{q_k} g_{k,i}=\widetilde{g}_j \quad \text{for all $i,j$},\\
& & & \text{such that $\P(J=j){I=i}=1$},
\end{aligned}
\end{equation}
for $k\in\{1,\ldots,K\}$. Similar to Problem~\eqref{eqn:N=1,epsilon>0,cases}, Problem~\eqref{eqn:cases} is maximizing a convex function over a convex bounded polytope. Therefore, we are able to solve Problem~\eqref{eqn:cases} by enumerating over the finite number of extreme points on the convex bounded polytope.

Next, for each $k\in\{1,\ldots,K\}$, we let $q_k^*(P_{J|I})$ be an optimal solution to Problem~\eqref{eqn:cases} corresponding to the conditional probability distribution $P_{J|I}$ and solve the following problem:
\begin{equation}\label{eqn:cases_2}
\begin{aligned}
P_{J|I}^*=&\underset{P_{J|I}}{\text{argmax}}
& &\sum_{k=1}^K \E[\KLD{T_{q_k^*(P_{J|I})} g_{k,I}}{T_{q_k^*(P_{J|I})} f_k}] \\
& \text{subject to}
& & \P(J=j){I=i}\in\{0,1\}\quad\text{for all $i,j$},\\
& & & \calL(I\to J)\leq \epsilon.
\end{aligned}
\end{equation}There is a maximum of $\stirling{|G|}{m}$ conditional distributions $P_{J|I}$ that satisfy $\calL(I\to J)\leq \epsilon$ where $m=\lfloor2^\epsilon\rfloor$ and $\stirling{a}{b}$ is the Stirling number of the second kind\cite{riordan2012introduction}. Thus, a solution for Problem~\eqref{eqn:cases_2} can be obtained by enumerating over the finite set of conditional probabilities $P_{J|I}$.

In the next proposition, we show that the set of sanitization channels obtained by the Local Exact method above is an optimal solution to Problem~\cref{eqn:optimize_relaxed_formulation} under the decentralized QCD setting.
\begin{Proposition}\label{prop:optimality}
Suppose the observations follow the signal model described in \eqref{eqn:iidsignalmodel}. Under the decentralized QCD setting, $\{q_1^*(P_{J|I}^*),\ldots,q_k^*(P_{J|I}^*)\}$ is an optimal solution to Problem~\cref{eqn:optimize_relaxed_formulation}. In particular, if $g_{1,i}=g_{2,i}=\ldots=g_{K,i}$ for $i\in\{1,\ldots,|G|\}$ then $\{q_1^*(P_{J|I}^*),\ldots,q_1^*(P_{J|I}^*)\}$ is an optimal solution to Problem~\cref{eqn:optimize_relaxed_formulation}.
\end{Proposition}
\begin{IEEEproof}
See \cref{sec:AppProp3}.
\end{IEEEproof}

\subsection{Sequential Hypothesis Testing Privacy}\label{subsec:decen_algorithms_sht}
Similar to the general case, Problem~\cref{eqn:optimize_formulation_asymptotic_relaxed_objective_sht} for the decentralized privacy-aware QCD problem is non-convex as the constraints $\calK_1(T_q)\leq \epsilon_1$ and $\calK_2(T_q)\geq \epsilon_2$ are non-convex. We use a restricted sanitization model to improve the computational tractability of Problem~\cref{eqn:optimize_formulation_asymptotic_relaxed_objective_sht}. However, in the decentralized version of the problem, each sensor is allowed to select its sanitization channel independent of the rest of the sensors. Let $\calC_k=\{q_{k,1},q_{k,2},\ldots,q_{k,n}\}$ be a finite set of local sanitization channels at sensor $k$ with column-stochastic matrices $T_{k,1},\ldots,T_{k,n}$. At each time instance $t$, we assume that sensor $k$ obtains an observation $Y_{k,t}=(Z_{k,t},A_{k,t})$ where $Z_{k,t}=q_{A_{k,t}}(X_{k,t})$ is a randomized function of the random variable $X_{k,t}$ under the sanitization channel $q_{A_{k,t}}$. We further assume that $\{A_{k,t}\}_{t\in\mathbb{N}}$ are i.i.d. generated with distribution $\phi_k$. Under the restricted sanitization model, rather than designing the sanitization channel $q$, we design the distribution $\phi_k$ that samples the sanitization channel $q_{A_{k,t}}$ from  $\calC_k$ at each sensor $k$ and time instance $t$. We then apply $q_{A_{k,t}}$ to $X_{k,t}$ to locally sanitize the signal.

Using similar arguments from \cite{lau2017optimal}, the asymptotic $\ARL$-$\EWADD$ trade-off of the GLR CuSum stopping time under the restricted sanitization model is given as
\begin{align*}
&\EWADD(\omega_{q})\\
&=\E[\frac{\log \gamma}{\sum_{k=1}^K\sum_{c=1}^n\phi_k(c)\KLD{T_{q_{k,c}} g_{k,I}}{T_{q_{k,c}} f_k}}](1+o(1)),
\end{align*}
as $\gamma\to\infty$. We also have
\begin{align*}
&\E[\KLD{T_{q}g_I}{T_{q}f}]\\
&=\E[\sum_{k=1}^K\sum_{c=1}^n\phi_k(c)\KLD{T_{q_{k,c}} g_{k,I}}{T_{q_{k,c}} f_k}].
\end{align*}
Thus, Problem~\cref{eqn:optimize_formulation_asymptotic_relaxed_objective_sht} becomes
\begin{equation}\label{eqn:optimize_formulation_asymptotic_relaxed_objective_sht_linearized_decent}
\begin{aligned}
& \underset{\phi}{\text{maximize}}
& & \sum_{k=1}^K\sum_{c=1}^n \phi_k(c)\E[\KLD{T_{q_{k,c}} g_{k,I}}{T_{q_{k,c}} f_k}] \\
& \text{subject to} &&\max_{i\in I_1} \min_{j\in I_1}\sum_{k=1}^K\sum_{c=1}^n\phi_k(c)\KLD{T_{q_{k,c}}g_{k,i}}{T_{q_{k,c}}g_{k,j}}\leq \epsilon_1,\\
& & &\sum_{k=1}^K\sum_{c=1}^n \phi_k(c)=1\\
& & &\phi_k(c)\geq 0, \quad\text{$c\in\{1,\ldots,n\}$ and $k\in\{1,\ldots,K\}$}\\
& & &\sum_{k=1}^K\sum_{c=1}^n\phi_k(c)\KLD{T_{q_{k,c}}g_{k,i}}{T_{q_{k,c}}g_{k,j}}\geq \epsilon_2\\
& & & \quad \text{for $i\in I_2$ and $j\in \{1,\ldots,|G|\}$}.
\end{aligned}
\end{equation}

Using similar arguments from \cref{prop:equiv_conditions}, Problem~\cref{eqn:optimize_formulation_asymptotic_relaxed_objective_sht_linearized_decent} is equivalent to the following MILP,
\begin{align}\label{eqn:optimize_formulation_asymptotic_relaxed_objective_sht_MILP_decent}
& \underset{\phi}{\text{maximize}}\ 
\sum_{k=1}^K\sum_{c=1}^n \phi_k(c)\E[\KLD{T_{q_{k,c}} g_{k,I}}{T_{q_{k,c}} f_k}] \\
& \text{subject to} \nn
& \xi(i)\leq \epsilon_1\quad\text{ for $i\in I_1$},\nn
& \delta(j,i)\in\{0,1\}\text{ for $i,j\in I_1$},\nn
& \sum_{k=1}^K\sum_{c=1}^n \phi_k(c)=1\nn
& \phi_k(c)\geq 0 \quad\text{$c\in\{1,\ldots,n\}$ and $k\in\{1,\ldots,K\}$} \nn
& \sum_{j\in I_1}\delta(j,i)=1\text{ for $i\in I_1$},\nn
& \sum_{k=1}^K\sum_{c=1}^n\phi_k(c)\KLD{T_{q_{k,c}}g_{k,i}}{T_{q_{k,c}}g_{k,j}}\nn
& \quad\quad\leq \xi(i)+(1-\delta(j,i))M\quad \text{for $i,j\in I_1$}.\nn
& \sum_{k=1}^K\sum_{c=1}^n\phi_k(c)\KLD{T_{q_{k,c}}g_{k,i}}{T_{q_{k,c}}g_{k,j}}\geq \epsilon_2\nn
& \quad \text{for $i\in I_2$ and $j\in \{1,\ldots,|G|\}$},\nn
& \sum_{k=1}^K\sum_{c=1}^n\phi_k(c)\KLD{T_{q_{k,c}}g_{k,i}}{T_{q_{k,c}}g_{k,j}}\geq \xi(i) \nn
& \quad\quad \text{for $i,j\in I_1$}.\nonumber
\end{align}
A global optimal solution to Problem~\cref{eqn:optimize_formulation_asymptotic_relaxed_objective_sht_MILP_decent} can be obtained using branch-and-bound methods on a linear program solver\cite{cvx}. 

\section{Numerical Experiments}\label{sec:numerical}
In this section, we present numerical results of experiments of the various methods introduced under the different signal models. The simulations were performed using MATLAB R2017a on a laptop with Intel(R) Core(TM) i7-6500U CPU@2.50GHz and 16.0GB RAM.
\subsection{Privacy-Aware QCD}
\subsubsection{Maximal Leakage Privacy}
\begin{figure}[!htbp]
\centering
\includegraphics[width=8cm]{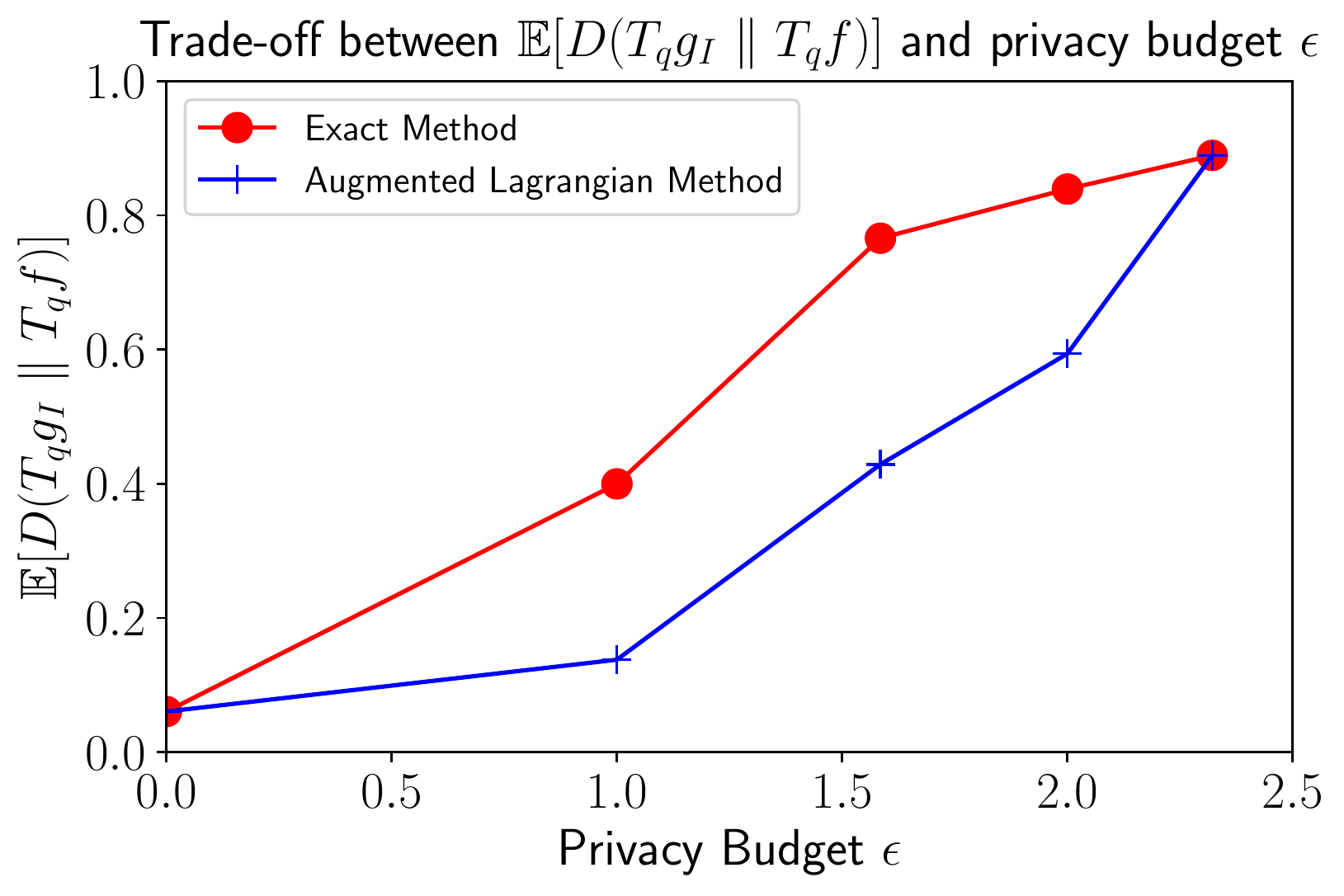}
\caption{Trade-off between the privacy budget $\epsilon$ and the expected KL divergence $\E[\KLD{T_{q}g_n}{T_{q}f}]$ for the Exact and augmented Lagrangian method.
}
\label{fig:compare_exact_al_method}
\end{figure}
In this subsection,  we present numerical results for the privacy-aware QCD task under the maximal leakage privacy metric. We consider the signal model with pre-change distribution $f$, set of possible post-change distribution $G=\{g_1,g_2,\ldots,g_5\}$ where $f$ and $G$ are randomly generated, uniform prior $p_I$ on the post-change distributions and $\calX=\calY=\{1,2,\ldots,7\}$. 

\begin{figure}[!htbp]
	\centering
	\includegraphics[width=8cm]{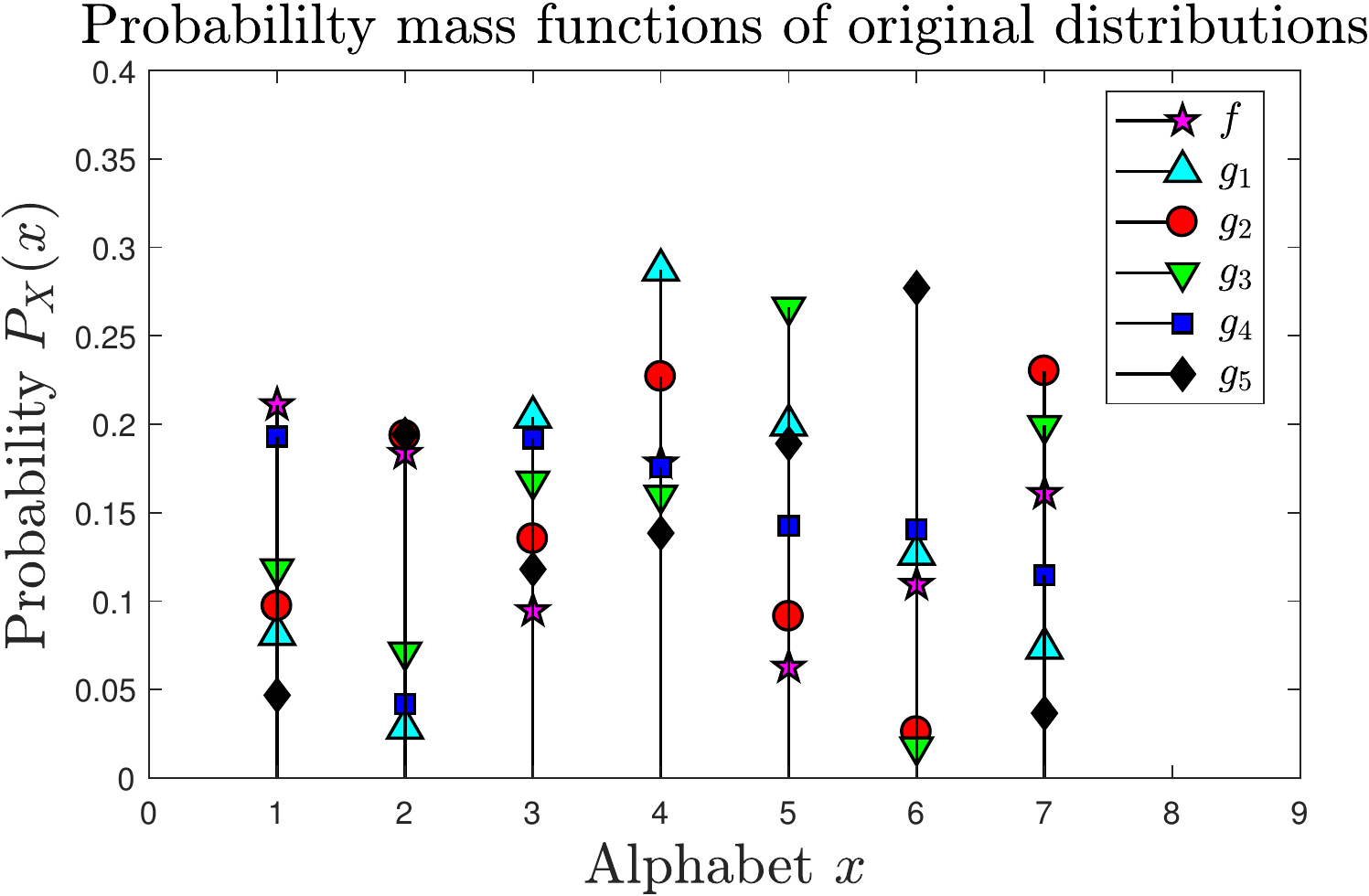}
	\caption{Pmf of pre- and post-change distributions.}
	\label{fig:sanitized5}
\end{figure}

\begin{figure}[!htbp]
	\centering
	\includegraphics[width=8cm]{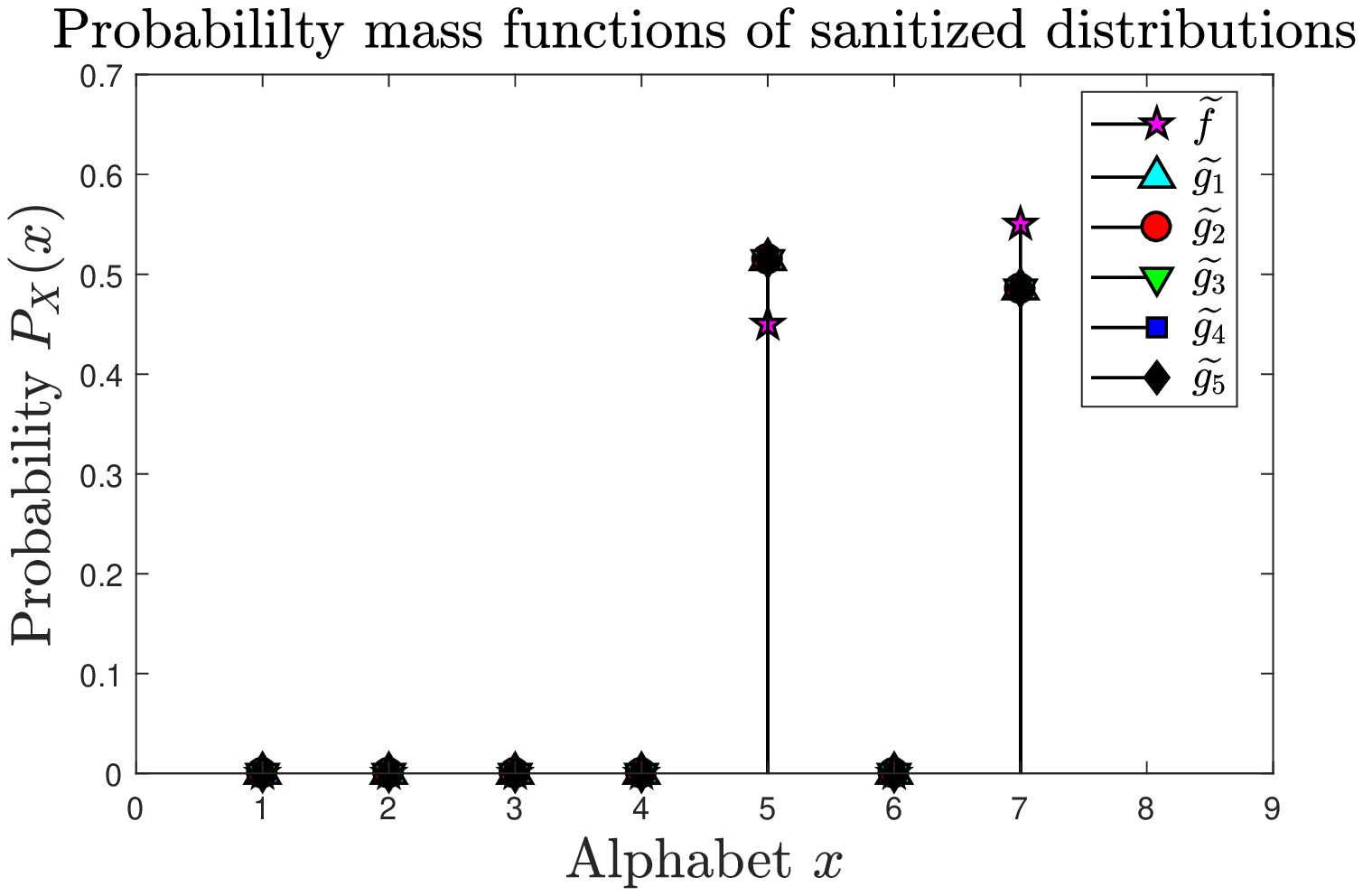}
	\caption{Pmf of pre- and post-change sanitized distributions for $\epsilon=\log_2 1$.}
	\label{fig:sanitized1}
\end{figure}

\begin{figure}[!htbp]
	\centering
	\includegraphics[width=8cm]{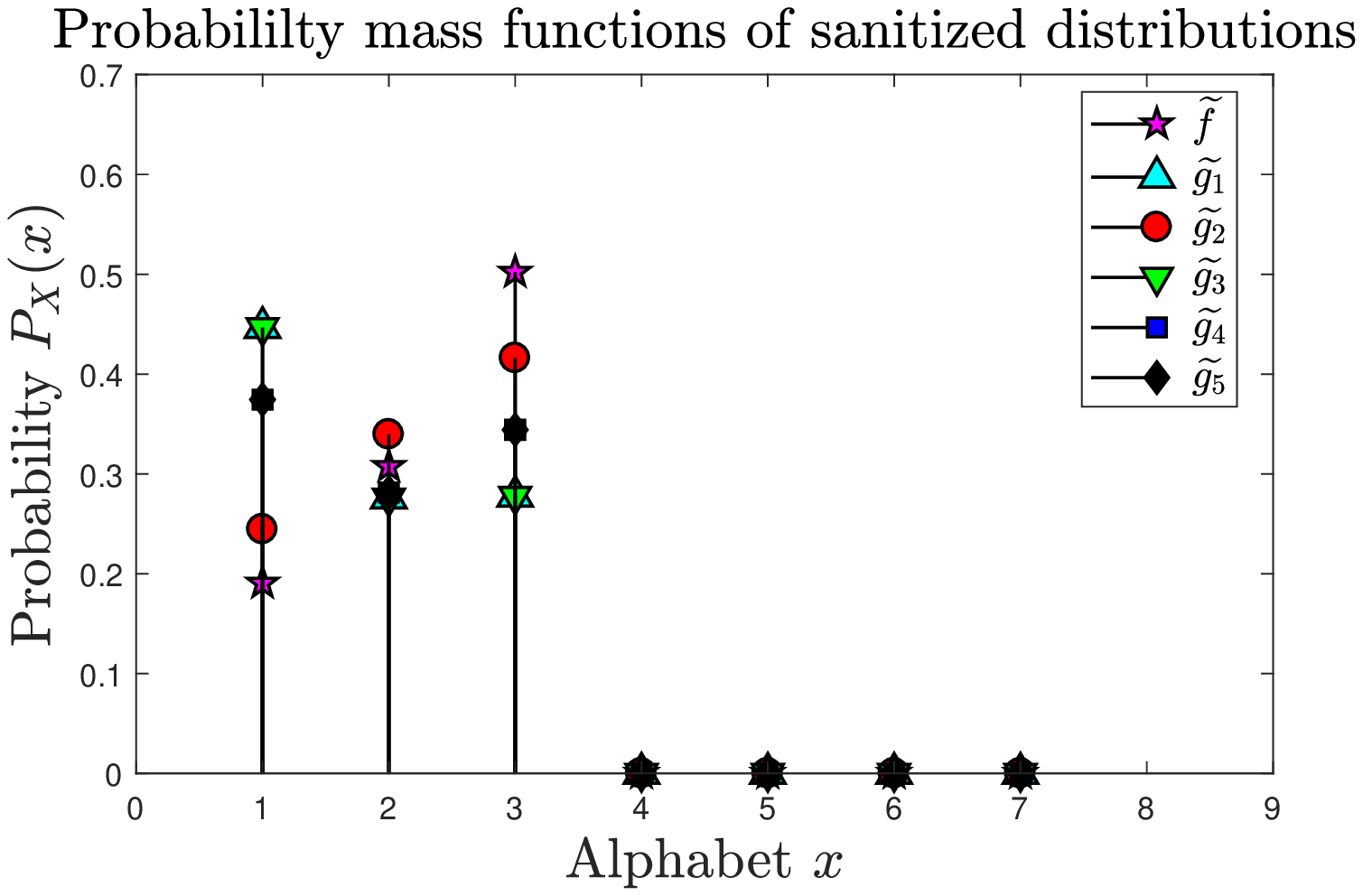}
	\caption{Pmf of pre- and post-change sanitized distributions for $\epsilon=\log_2 3$.}
	\label{fig:sanitized3}
\end{figure}

\begin{figure}[!htbp]
\centering
\includegraphics[width=8cm]{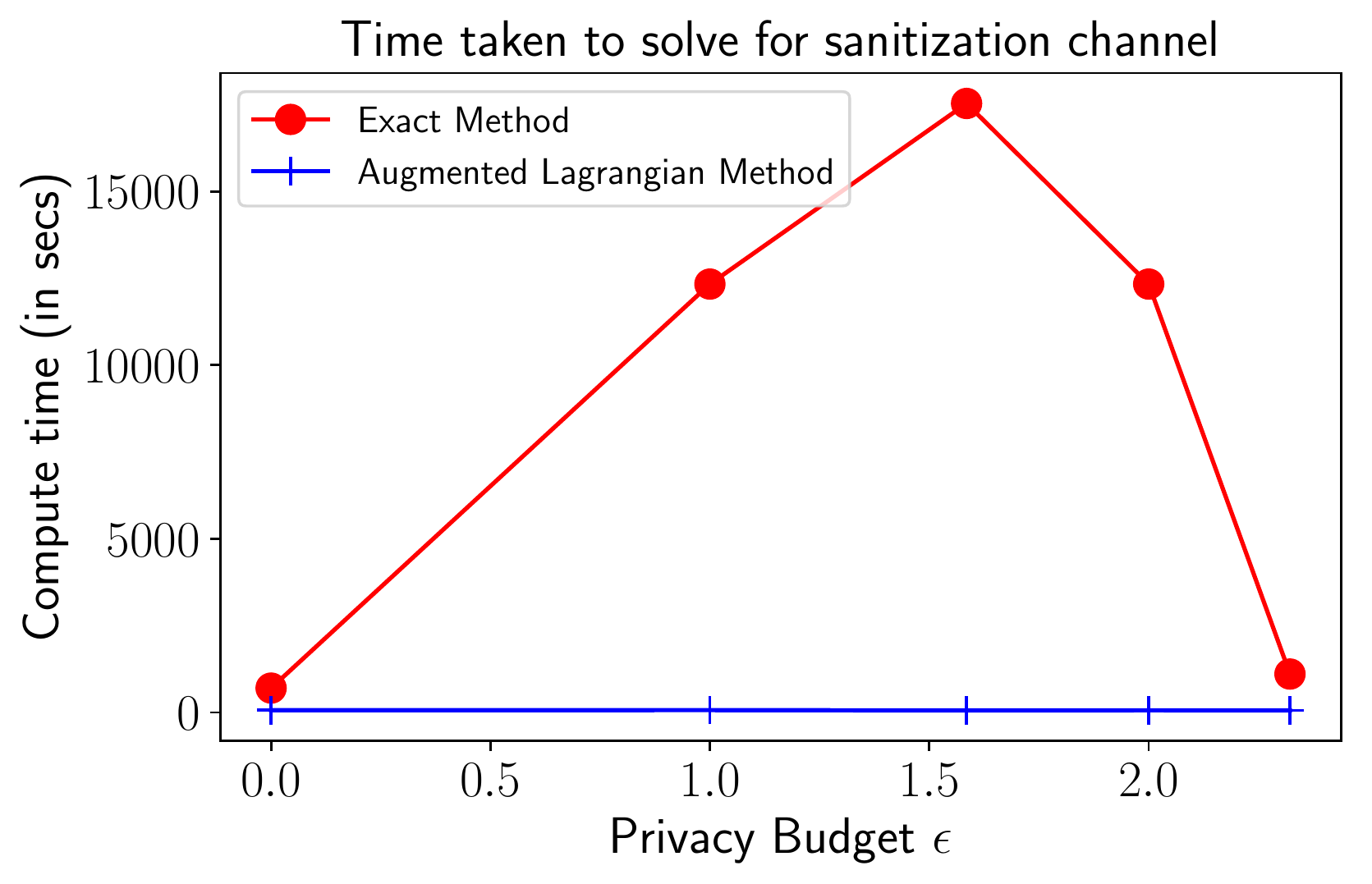}
\caption{Comparison of the time taken to solve the relaxed channel design problem for the Exact and augmented Lagrangian method.}
\label{fig:compare_exact_al_method_time}
\end{figure}

Using algorithms described in \cref{subsec:algorithms_ml}, we solve the relaxed channel design problem~\cref{eqn:optimize_relaxed_formulation} exactly and problem \cref{eqn:optimize_continuous_relaxed_formulation} using the augmented Lagrangian method. First, we present results to illustrate  the trade-off between the privacy budget $\epsilon$ and the expected KL divergence. In \cref{fig:sanitized5,fig:sanitized1,fig:sanitized3}, we plot the pmf of the original distributions, sanitized distributions when $\epsilon=\log_2 1$ and sanitized distributions when $\epsilon=\log_2 3$, respectively. In \figref{fig:compare_exact_al_method}, the expected KL divergence obtained by each of the methods is plotted against the privacy budget $\epsilon$. When the privacy budget $\epsilon=\log_2 5\approx 2.3$, the privacy constraint $\calL_{\text{max}}(I\to J)\leq \epsilon$ becomes redundant as it is trivially satisfied. Problem~\eqref{eqn:optimize_formulation} then reduces to a standard QCD problem without any sanitization of the observations. It should be noted that two graphs intersect the y-axis at a positive value rather than at zero. This is because when our privacy budget $\epsilon$ is zero, we only remove all information that allows us to identify the post-change distributions. Thus, it is possible to distinguish the post-change distributions from the pre-change distribution in some cases. As the augmented Lagrangian method only guarantees local optimality, the channel obtained by solving Problem \cref{eqn:optimize_continuous_relaxed_formulation} achieves a lower value of the objective function as compared to the channel obtained by solving  Problem~\cref{eqn:optimize_relaxed_formulation}. Next, we plot the compute time required for each of the methods against the privacy budget $\epsilon$ in \figref{fig:compare_exact_al_method_time}. The results indicate that the augmented Lagrangian method requires significantly lesser compute time compared to the Exact method. It should be noted that the compute time required by the Exact Method is directly proportional to $\stirling{|G|}{m}$. The peak observed in \figref{fig:compare_exact_al_method_time} corresponds to $\stirling{|G|}{m}$ achieving its maximum at $m=3$ when $|G|=5$. 


\subsubsection{Sequential Hypothesis Testing Privacy}
In this subsection,  we present numerical results for the centralized QCD task under the sequential hypothesis testing privacy metric. We consider the signal model with pre-change distribution $f$, set of possible post-change distribution $G=\{g_1,g_2,\ldots,g_6\}$ with $I_1=\{1,2,3\}$ and $I_2=\{4,5,6\}$, a uniform prior $p_I$ on the post-change distributions, $\calX=\calY=\{1,2,3,4\}$ and the set of deterministic functions from $\calX$ to $\calY$ as the finite set of sanitization channel $\calC$. We generate $f$ and $G$ randomly. 

\begin{figure}[!htbp]
	\centering
	\includegraphics[width=8cm]{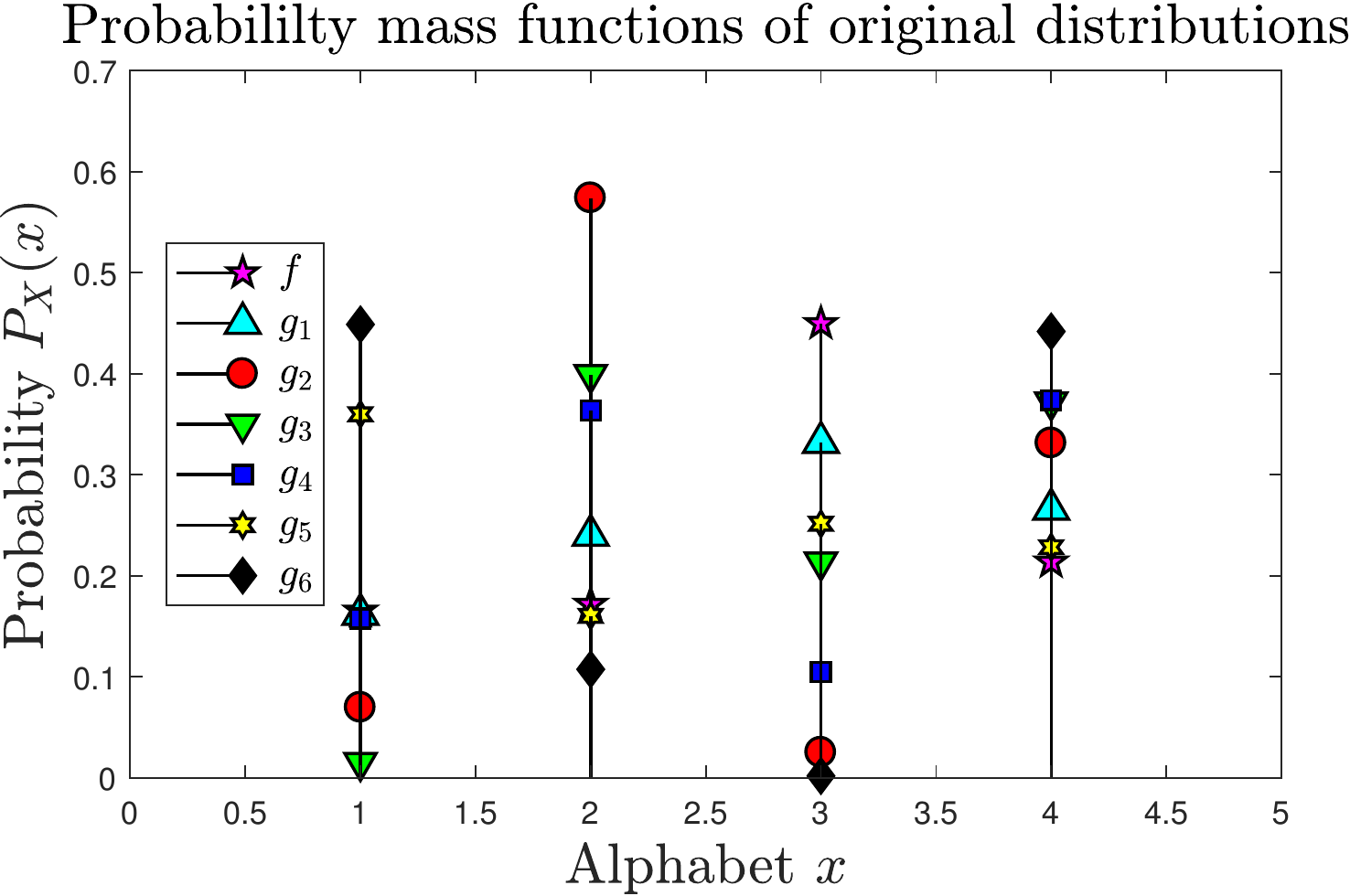}
	\caption{Pmf of pre- and post-change original distributions.}
	\label{fig:original_seq}
\end{figure}

\begin{figure}[!htbp]
	\centering
	\includegraphics[width=8cm]{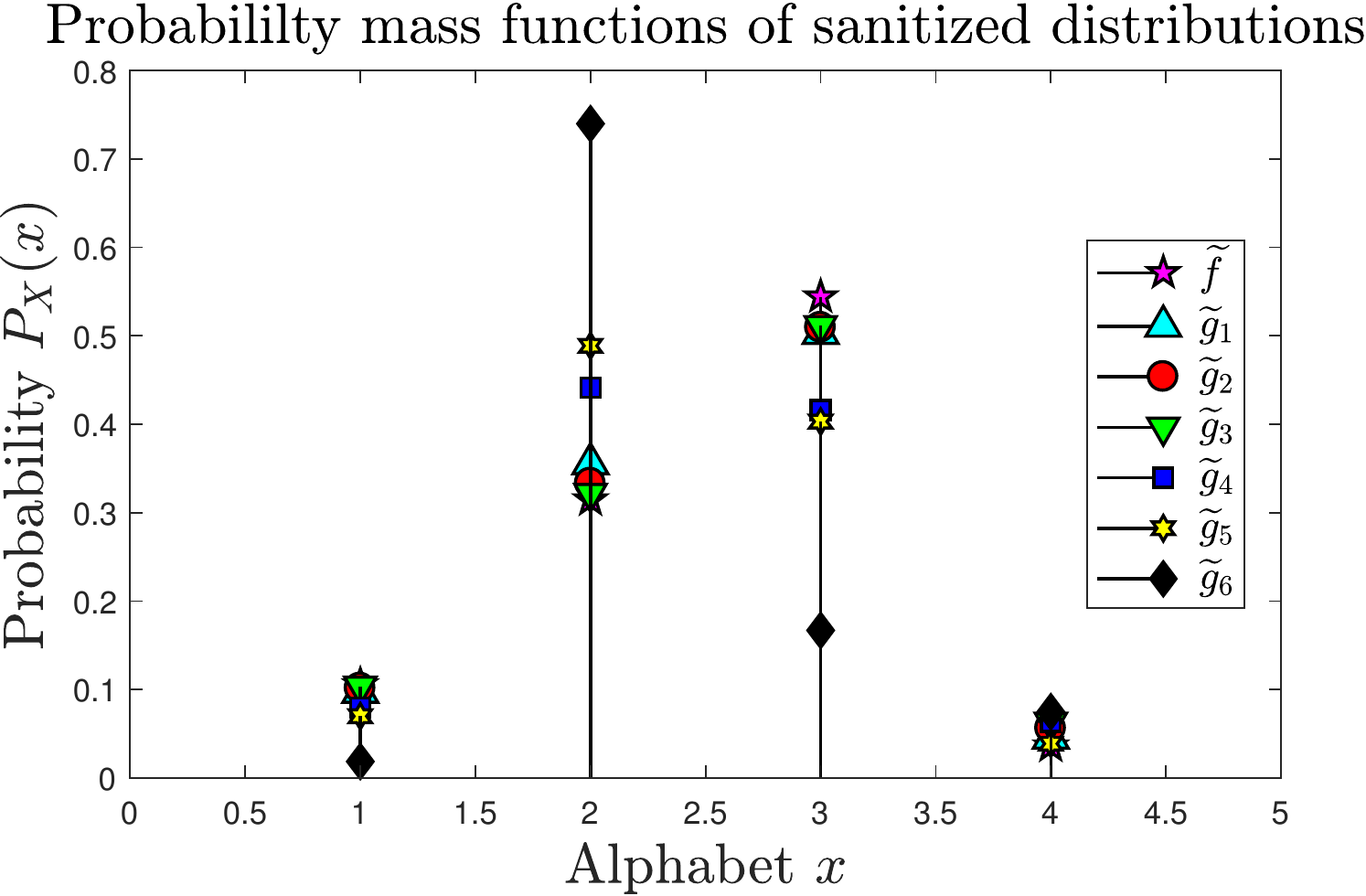}
	\caption{Pmf of pre- and post-change sanitized distributions for $\epsilon_1=0.0012$ and $\epsilon_2=0.0024$.}
	\label{fig:sanitized_seq}
\end{figure}

\begin{figure}[!htbp]
\centering
\includegraphics[width=8cm]{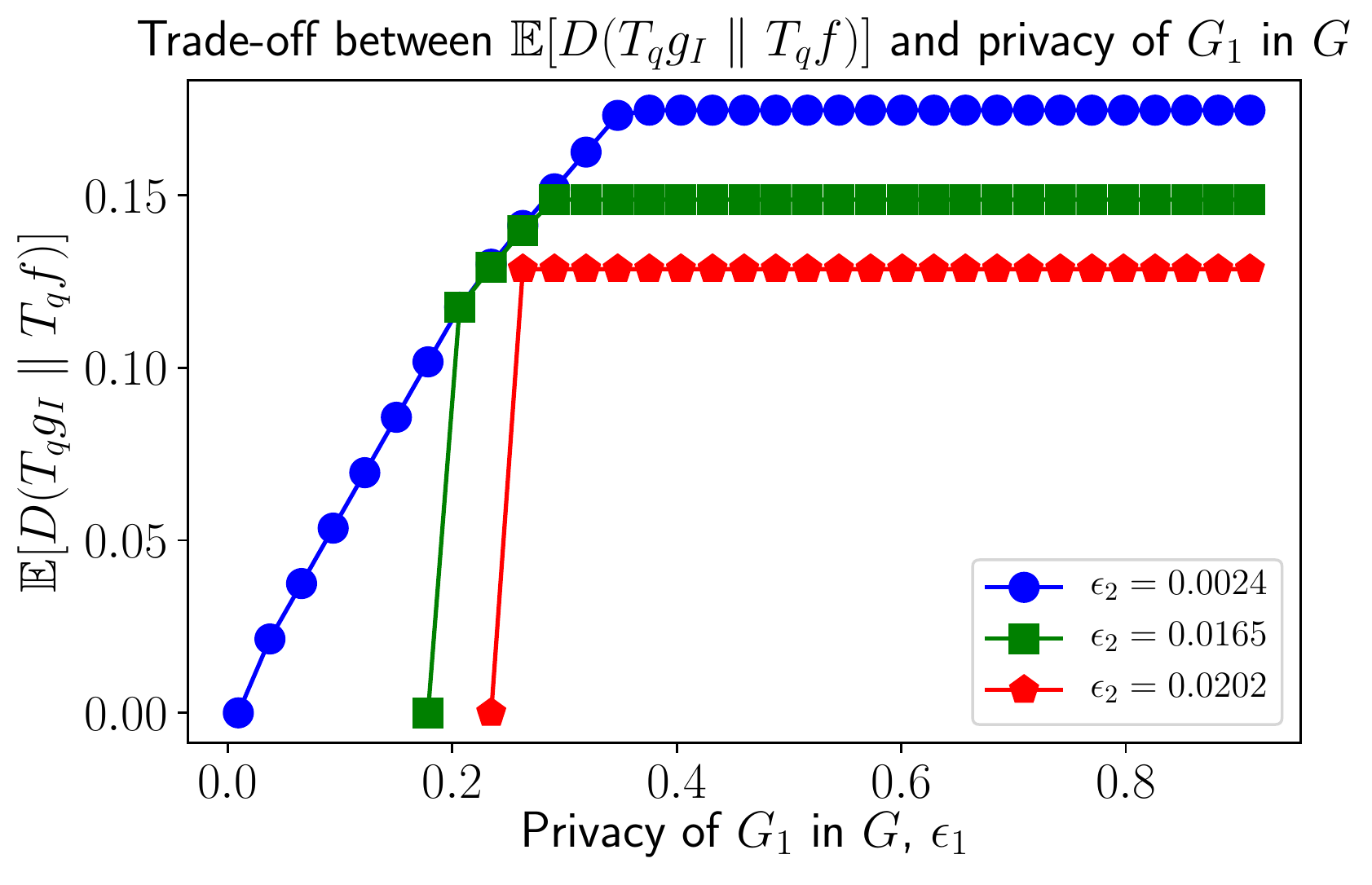}
\caption{Trade-off between the privacy budget $\epsilon_1$ and the expected KL divergence for different distinguishability levels $\epsilon_2$.}
\label{fig:compare_privacy_vs_kl}
\end{figure}

\begin{figure}[!htbp]
\centering
\includegraphics[width=8cm]{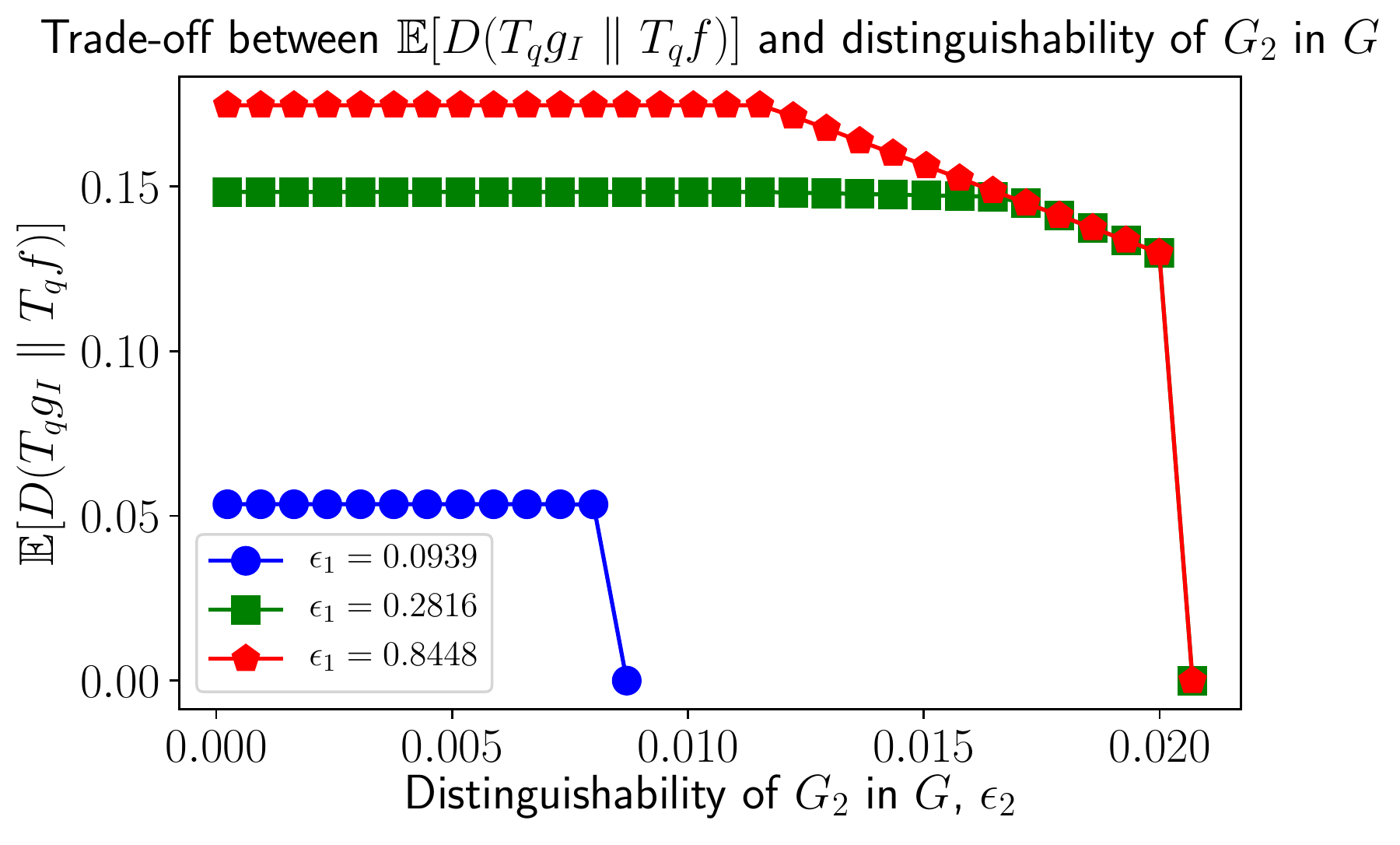}
\caption{Comparison of trade-off between the distinguishability $\epsilon_2$ of $I_2$ in $G$ and the expected KL divergence for different privacy $\epsilon_1$ of $I_1$ in $G$.}
\label{fig:compare_distinguishability_vs_kl}
\end{figure}

Using algorithms described in \cref{subsec:algorithms_sht}, we solve the relaxed channel design problem~\cref{eqn:optimize_formulation_asymptotic_relaxed_objective_sht_linearized} by solving the MILP in Problem~\cref{eqn:optimize_formulation_asymptotic_relaxed_objective_sht_MILP}. By the data processing inequality $\calK_1(T_q)\leq \max_{i\in I_1}\min_{j\in I_1}\KLD{g_i}{g_j}=0.9387$ and $\calK_2(T_q)\leq \min_{i\in I_2}\min_{j\in I_1\cup I_2}\KLD{g_i}{g_j}=0.0855$, thus we focus our attention to the region where $0\leq \epsilon_1\leq 0.9387 $ and $0\leq\epsilon_2\leq 0.0855$. In \cref{fig:original_seq,fig:sanitized_seq}, we plot the pmf of the original distributions, sanitized distributions when $\epsilon_1=0.0012, \epsilon_2=0.0024$, respectively. We compare the trade-off between the privacy $\epsilon_1$ of $I_1$ in $G$ and the expected KL divergence in \figref{fig:compare_privacy_vs_kl} and the trade-off between the distinguishability $\epsilon_2$ of $I_1$ in $G$ and the expected KL divergence in \figref{fig:compare_distinguishability_vs_kl}. In both \figref{fig:compare_privacy_vs_kl} and \figref{fig:compare_distinguishability_vs_kl}, we observe that the average KL divergence increases as $\epsilon_1$ increases and $\epsilon_2$ decreases which is consistent with the behaviour expected of the optimal value of Problem~\cref{eqn:optimize_formulation_asymptotic_relaxed_objective_sht_linearized}.


\subsection{Decentralized Privacy-Aware QCD}

\subsubsection{Maximal Leakage Privacy}
In this subsection,  we present numerical results for the decentralized QCD task under the maximal leakage privacy metric. We consider the signal model described in \eqref{eqn:iidsignalmodel}. We generate $f_k$ and $g_{k,i}$ randomly such that $g_{k_1,i}=g_{k_2,i}$ for all $k_1,k_2\in\{1,\ldots,K\}$ and $i\in\{1,\ldots,5\}$, use a uniform prior $p_I$ on the post-change distributions and let $\calX=\calY=\{1,2,\ldots,5\}$.
\begin{figure}[!htbp]
\centering
\includegraphics[width=8cm]{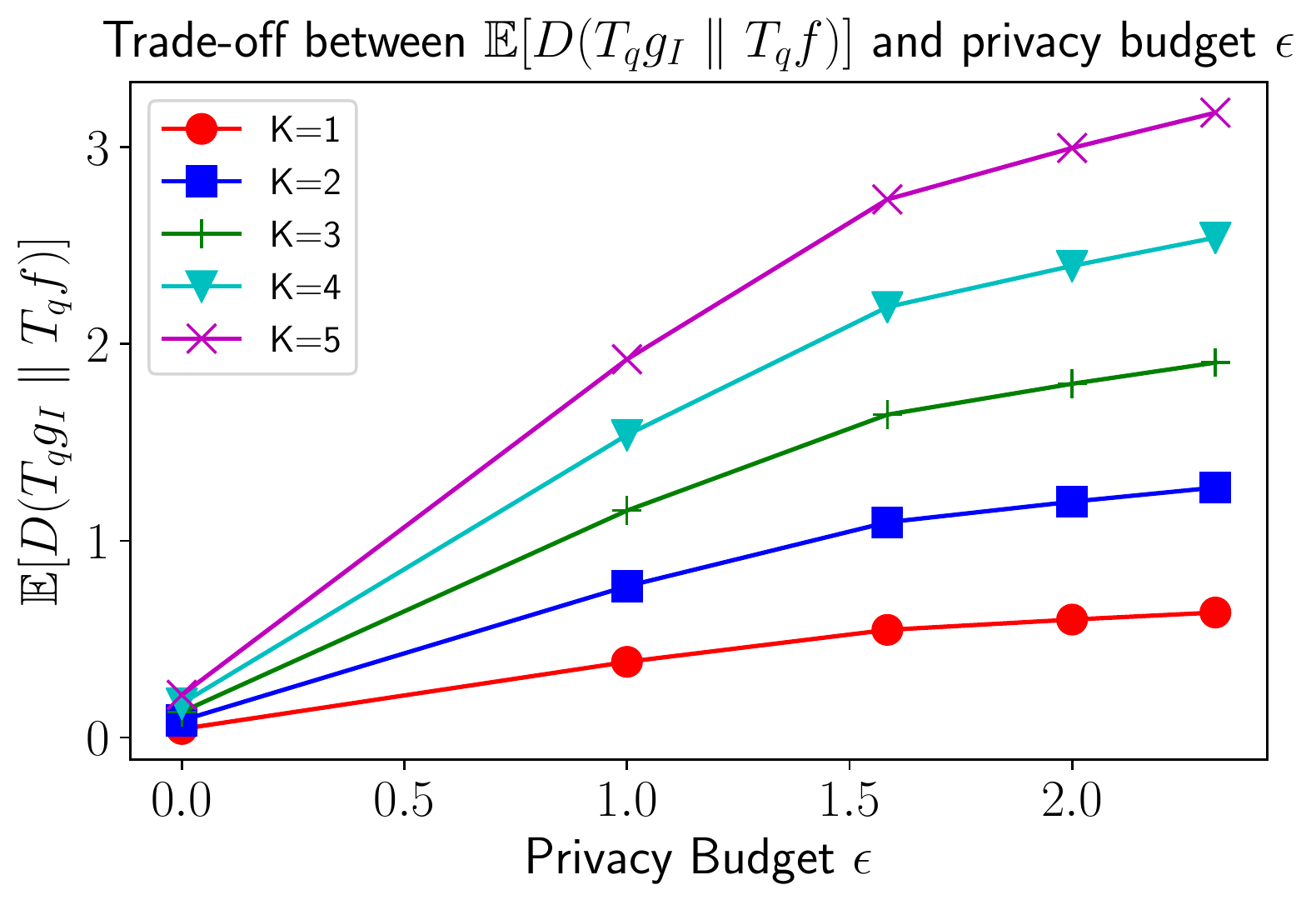}
\caption{The trade-off between the expected KL divergence and the privacy budget $\epsilon$ for different number of sensors.}
\label{fig:privacy_util_tradeoff_decent}
\end{figure}
\begin{figure}[!htbp]
\centering
\includegraphics[width=8cm]{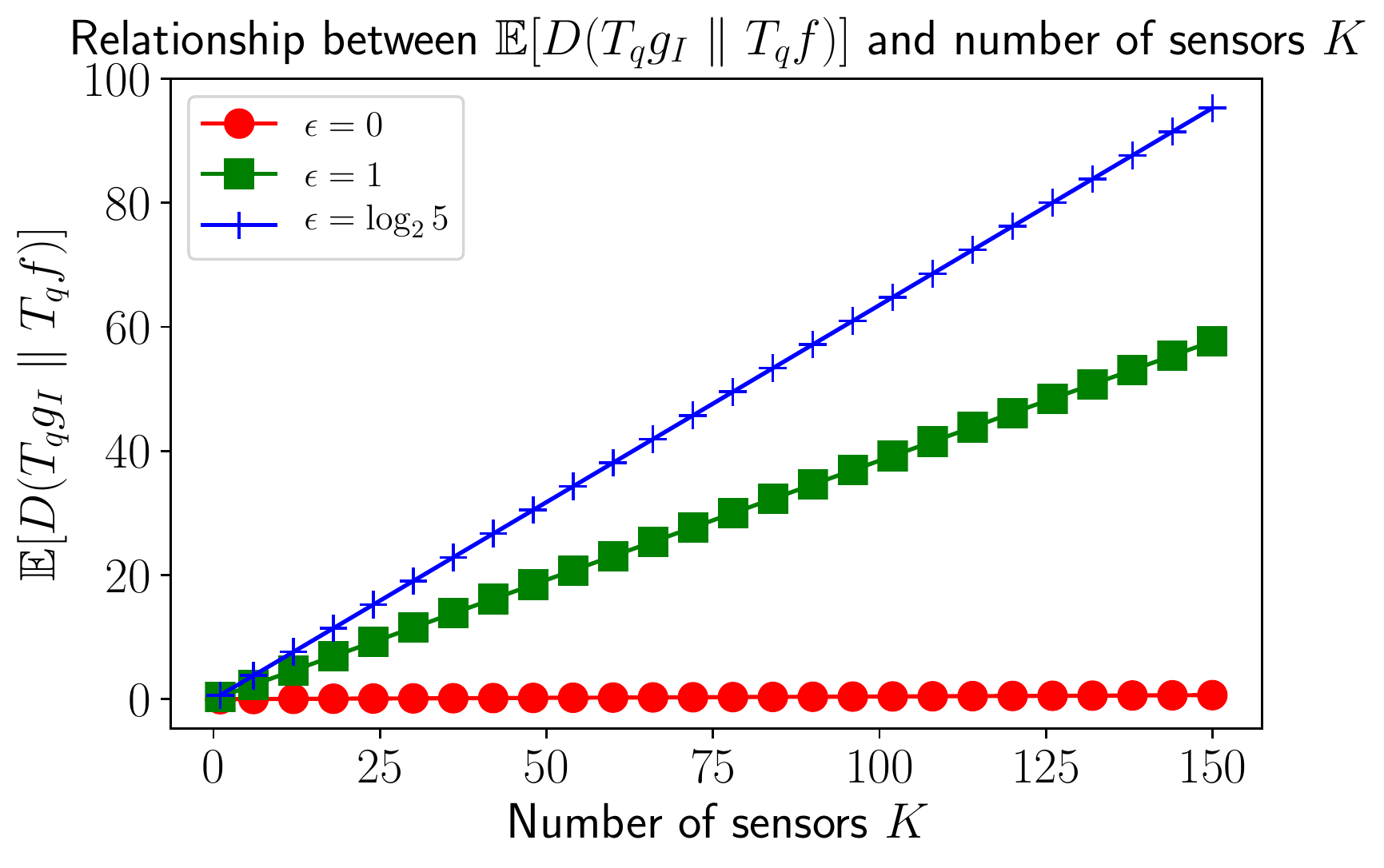}
\caption{The expected KL divergence achieved by the solved sanitization channel for different number of sensors.}
\label{fig:privacy_util_tradeoff_decent_linear}
\end{figure}

We solve the relaxed channel design problem~\cref{eqn:optimize_relaxed_formulation} using the Local Exact method described in \cref{subsec:decen_algorithms_ml} for different number of sensors $K$. In \figref{fig:privacy_util_tradeoff_decent}, we plot the trade-off between the expected KL divergence and the privacy budget $\epsilon$ for different number of sensors $K$. In \figref{fig:privacy_util_tradeoff_decent_linear}, we plot the expected KL divergence for the privacy budget $\epsilon=0,1,\log_2 5$ as the number of sensors vary between $1$ and $150$. The results from the experiments indicate that for the case where the signal observed at each of sensors are identically distributed, the expected KL divergence grows linearly with respect the number of sensors $K$. Next, we compare the compute time taken to solve Problem~\eqref{eqn:optimize_relaxed_formulation} using the Exact method and the Local exact method. We solve the relaxed channel design problem~\cref{eqn:optimize_relaxed_formulation} with a privacy budget of $\epsilon=1$ using the Exact method described in \cref{subsec:algorithms_ml} and the Local Exact method described in \cref{sec:iid}. \begin{figure}[!htbp]
\centering
\includegraphics[width=8cm]{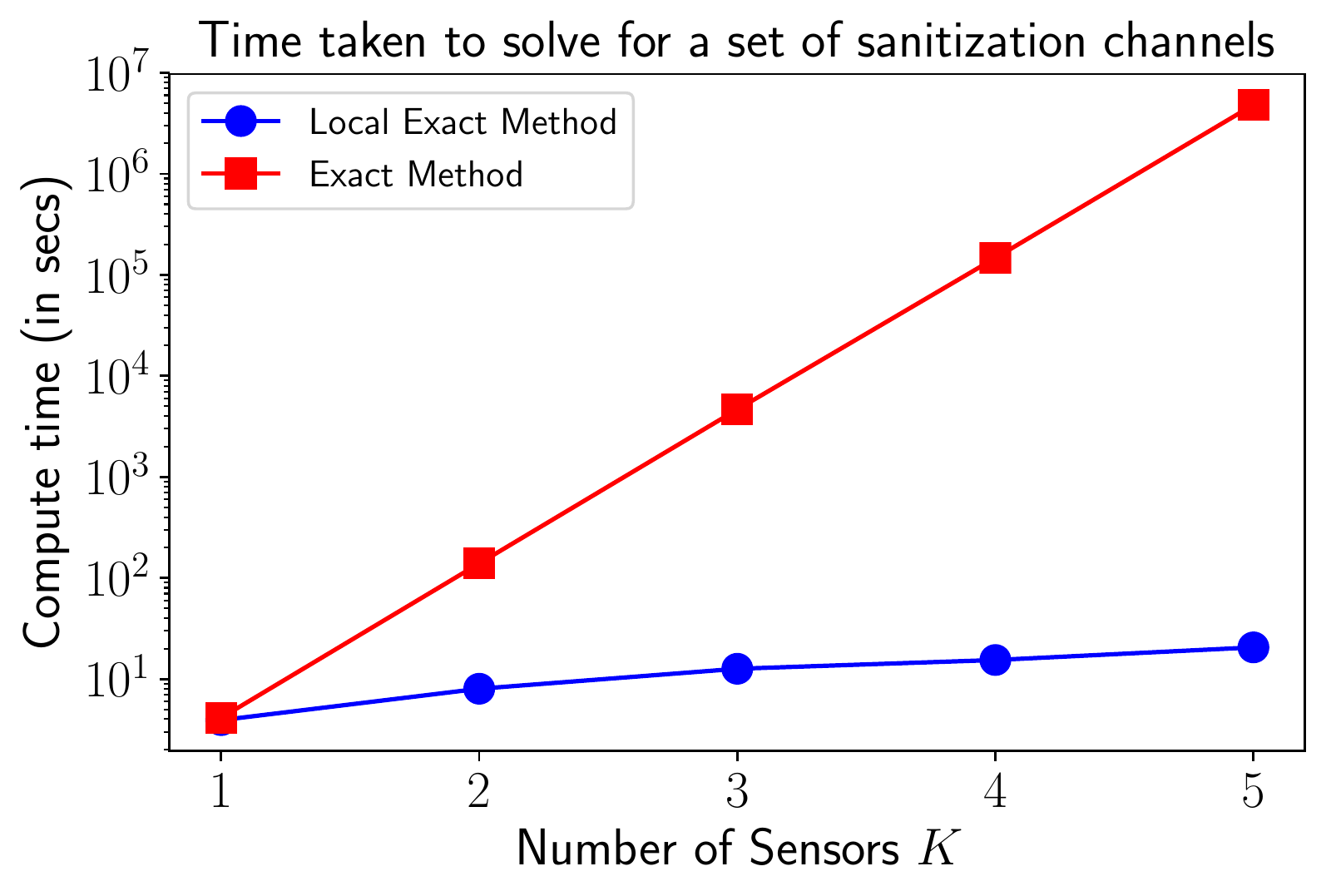}
\caption{Comparison of compute time for the Exact and Local Exact method.}
\label{fig:compare_exact_method}
\end{figure}In \figref{fig:compare_exact_method}, we  plot the compute time required to solve Problem~\eqref{eqn:optimize_relaxed_formulation} against the number of sensors $K$ for the Exact method and the Local exact method respective. The simulation result indication that the compute time required by Exact method increases exponentially with respect to the number of sensors $K$ while the compute time required by the Local Exact method remains reasonably low.

\subsubsection{Sequential Hypothesis Testing Privacy}
In this subsection,  we present numerical results for the decentralized QCD task under the sequential hypothesis testing privacy metric. We consider the signal model described in \eqref{eqn:iidsignalmodel}. We generate $f_k$ and $g_{k,i}$ randomly such that $g_{k_1,i}=g_{k_2,i}$ for all $k_1,k_2\in\{1,\ldots,K\}$ and $i\in\{1,\ldots,6\}$, use a uniform prior $p_I$ on the post-change distributions, partition the index set of $G$ into $I_1=\{1,2,3,4\}$ and $I_2=\{5,6\}$, and let $\calX=\calY=\{1,2,3\}$. 
\begin{figure}[!htbp]
\centering
\includegraphics[width=8cm]{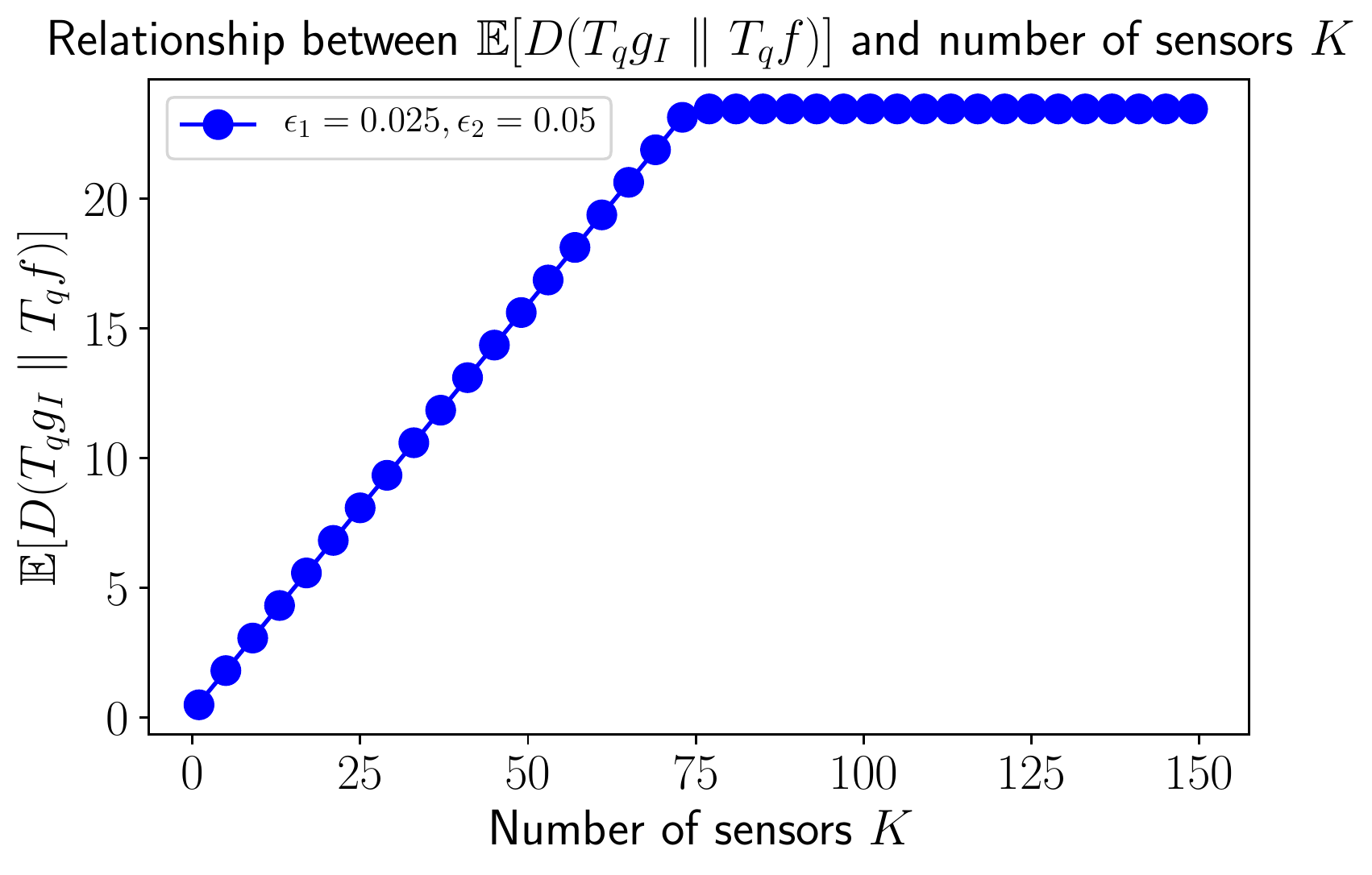}
\caption{The trade-off between the expected KL divergence and the privacy budget $\epsilon$ for different number of sensors.}
\label{fig:privacy_util_tradeoff_decent_sht}
\end{figure}
We solve the relaxed channel design problem~\cref{eqn:optimize_relaxed_formulation} using by solving the MILP described in \cref{subsec:decen_algorithms_sht} for the privacy budget $\epsilon_1=0.025,\epsilon_2=0.05$. In \figref{fig:privacy_util_tradeoff_decent_sht}, we present the simulation results to illustrate the relationship between the number of sensors and the expected KL divergence under the decentralized signal model. As the number of sensor increases, the privacy constraint $\calK_1(T_q)\leq\epsilon_1$ becomes more difficult to satisfy and the distinguishably constraint $\calK_2(T_q)\geq \epsilon_2$ becomes easier to satisfy. Thus, we do not expect that the expected KL divergence to grow linearly with the number of sensors. The simulation indicates that the expected KL divegerence does not increase linearly with the number of sensors $K$.

\begin{figure}[!htbp]
\centering
\includegraphics[width=8cm]{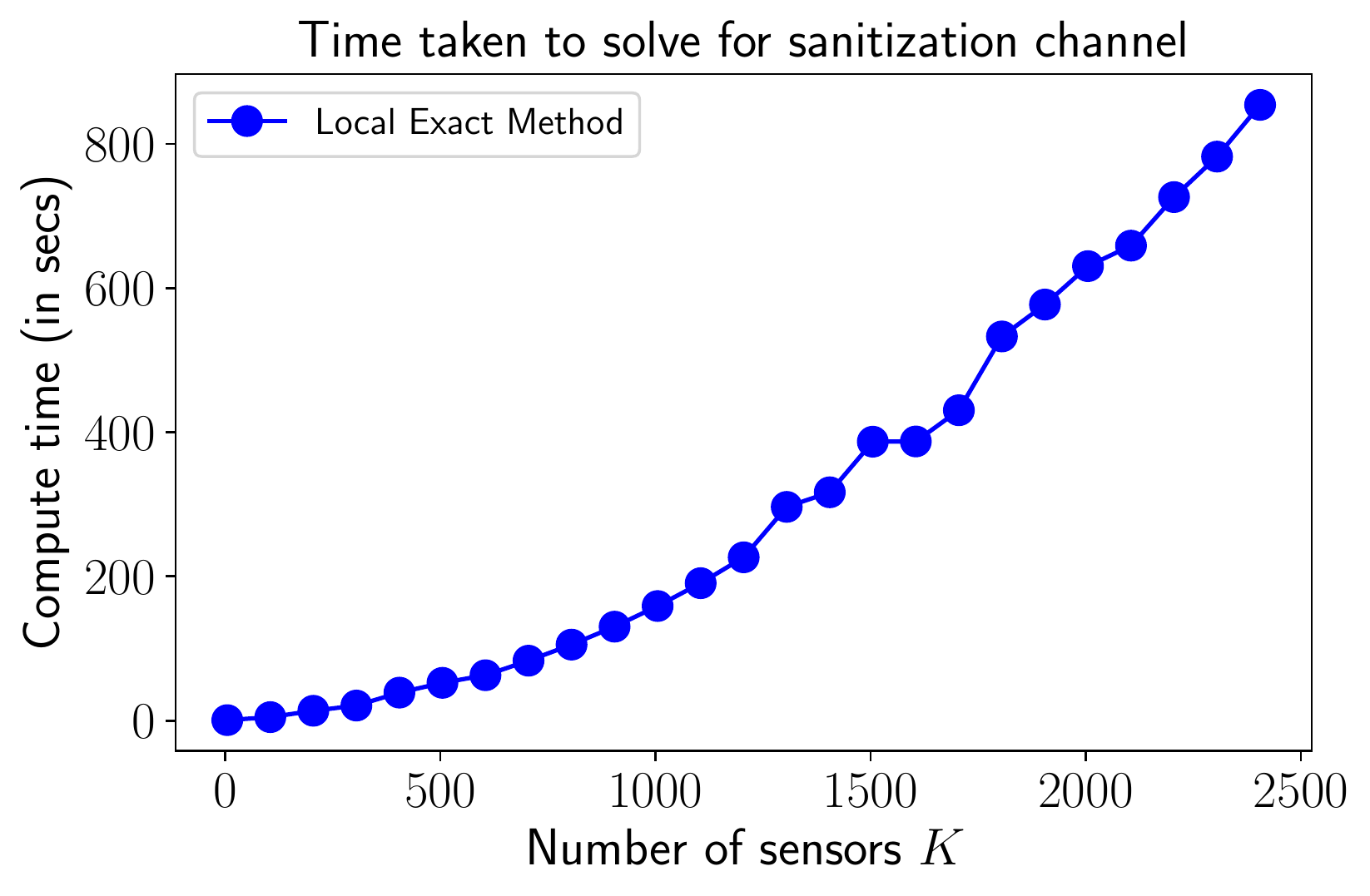}
\caption{Relationship between average compute time and the number of sensors $K$ .}
\label{fig:compare_seq_hypo_time}
\end{figure}

Next, we present the relationship between the average compute time required to solve Problem~\eqref{eqn:optimize_relaxed_formulation} and the number of sensors $K$. We randomly generate 500 sets of distributions satisfying $g_{k_1,i}=g_{k_2,i}$ for all $k_1,k_2\in\{1,\ldots,K\}$ and $i\in\{1,\ldots,6\}$. For each set of distribution, we randomly sample $\epsilon_1$ with uniform probability in the interval $[0,\max_{i\in I_1}\min_{j\in I_1}\KLD{g_i}{g_j}]$ and $\epsilon_2$ with uniform probability in the interval $[0,\min_{i\in I_2}\min_{j\in I_1\cup I_2}\KLD{g_i}{g_j}]$. In the event when the sampled $\epsilon_1,\epsilon_2$ makes  Problem~\eqref{eqn:optimize_relaxed_formulation} infeasible, we resample $\epsilon_1,\epsilon_2$. The compute time taken to solve the MILP for each set of distributions is recorded and the relationship between the average compute time and the number of sensors is present in \figref{fig:compare_seq_hypo_time}. The simulation results suggests that the average compute time increases with respest to the number of sensors $K$ in a super-linear manner. This may be a potential challenge for applications when the number of sensors is large. 

\section{Conclusion and future work}\label{sec:conclusion}
In this paper, we have proposed a framework for privacy-aware QCD using two different privacy metrics for both centralized and decentralized QCD tasks. We also proposed optimization problems for which the solution provides a sanitization channel where the GLR CuSum stopping time is asymptotically optimal under each of the scenarios. We derived relaxations to the channel design problem and provided algorithms to obtain exact solutions to the relaxed channel design problem. For the maximal leakage privacy metric, a continuous relaxation for the channel design problem is proposed so that locally optimal solutions can be obtained when the exact solutions are computationally intractable. An algorithm that scales linearly with the number of sensors is also proposed for the decentralized QCD tasks when the signal recieved at the sensors are mutually independent. One drawback of the proposed signal and sanitization model for privacy-aware QCD under maximal leakage privacy metric is the discreteness of the privacy constraint $\calL_{\text{max}}(I\to J)$. Using a randomization of multiple possible sanitization channels, the privacy constraint $\calL_{\text{max}}(I\to J)$ is able to achieve a continuous interval of values. The optimization problem related to this new channel design problem is, however, much harder to solve and is a potential direction for future work. For the sequential hypothesis testing privacy metric, a modified sanitization model is proposed to relax the channel design problem into a MILP for both centralized and decentralized QCD tasks. One interesting direction of future work can be done to determine the pairs $(\epsilon_1,\epsilon_2)$ where Problem~\cref{eqn:optimize_formulation_asymptotic_relaxed_objective_sht_linearized} is feasible. It will also be interesting to design an algorithm with computation complexity that grows sub-linearly with the number of sensors for the decentralized QCD task with the sequential hypothesis testing privacy metric.

\appendices

\section{Proof of \cref{prop:upperbound_max_leakage}}\label[appendix]{sec:AppProp1}


We define the following matrices
$\Delta\in\mathbb{R}^{G\times\calU}$, $\Theta\in\mathbb{R}^{\calY^{t-\nu+1}\times\widetilde{G}}$, $\Lambda\in\mathbb{R}^{\widetilde{G}\times G}$, $\Phi_0\in\mathbb{R}^{U\times\calY^{t-\nu+1}}$ and $\Phi_1\in\mathbb{R}^{U\times\widetilde{G}}$ such that 
\begin{align}\label{eqn:matrices_1}
&[\Delta]_{i,u}= P_{U|I}(u\ |\ i)P_I(i),\nn
&[\Theta]_{y^{\nu:t},j}=P_{Y^{\nu:t}|J}(y^{\nu:t}\ |\ j),\\
&[\Lambda]_{j,i}==P_{J|I}(j\ |\ i),\nonumber
\end{align}
and
\begin{align}\label{eqn:matrices_2}
[\Phi_0]_{u,y^{\nu:t}}&=\begin{cases}
1\quad\text{if $u=\argmax_{u\in\calU}P_{\calU,Y^{\nu:t}}(u,y^{\nu:t})$,}\\
0\quad\text{otherwise.}
\end{cases}\\
[\Phi_1]_{u,j}&=\begin{cases}
1\quad\text{if $u=\argmax_{u\in\calU}P_{\calU,J}(u,j)$,}\\
0\quad\text{otherwise.}
\end{cases}\nonumber
\end{align}
For a fixed random variable $U$ that is a randomized function of $I$, we have
\begin{align}\label{eqn:demoninator_ml_1}
&\sum_{y^{\nu:t}}\max_{u\in\calU}P_{\calU,Y^{\nu:t}}(u,y^{\nu:t})\nonumber\\
&=\sum_{y^{\nu:t}}\max_{u\in\calU}\sum_{j=1}^{|\widetilde{G}|}\sum_{i=1}^{|G|}P_{J,I,Y^{\nu:t},\calU}(j,i,y^{\nu:t},u)\nonumber\\
&=\sum_{y^{\nu:t}}\max_{u\in\calU}\sum_{i=1}^{|G|}P_{\calU|I}(u\ |\ i)P_I(i)\nonumber\\  &\quad\quad\quad\left(\sum_{j=1}^{|\widetilde{G}|}P_{Y^{\nu:t}|J}(y^{\nu:t}\ |\ j)P_{J|I}(j\ |\ i)\right).
\end{align}
Similarly, we have
\begin{align}\label{eqn:demoninator_ml_2}
\sum_{j=1}^{|\widetilde{G}|}\max_{u\in\calU}P_{\calU,J}(u,j)&=\sum_{j=1}^{|\widetilde{G}|}\max_{u\in\calU}\sum_{y^{\nu:t}}\sum_{i=1}^{|G|}P_{J,I,Y^{\nu:t},\calU}(j,i,y^{\nu:t},u)\nonumber\\
&=\sum_{j=1}^{|\widetilde{G}|}\max_{u\in\calU}\sum_{i=1}^{|G|}P_{\calU|I}(u\ |\ i)P_{I}(i) P_{J|I}(j\ |\ i).
\end{align}
From \cref{eqn:matrices_1,eqn:matrices_2}, we can see that \cref{eqn:demoninator_ml_1,eqn:demoninator_ml_2} can be expressed using matrix operations
\begin{align*}
&\sum_{j=1}^{|\widetilde{G}|}\max_{u\in\calU}P_{\calU,J}(u,j)=\text{Trace}\left(\Lambda\Delta\Phi_1\right),\\
&\sum_{y^{\nu:t}}\max_{u\in\calU}P_{\calU,Y^{\nu:t}}(u,y^{\nu:t})=\text{Trace}\left(\Theta\Lambda\Delta\Phi_0\right),
\end{align*}
Furthermore, $\Phi_1$ is a solution to the following problem,
\begin{align*}
\Phi_1=\argmax_{\Gamma}\text{Trace}(\Lambda\Delta\Gamma),
\end{align*}
where the maximization is taken over all column stochastic matrices $\Gamma$. Since $\Phi_0,\Theta$ are column stochastic matrices, we have
\begin{align*}
\text{Trace}\left(\Lambda\Delta\Phi_0\Theta\right)\leq \text{Trace}\left(\Lambda\Delta\Phi_1\right),
\end{align*}
and thus \begin{align*}
\sum_{y^{\nu:t}}\max_{u\in\calU}P_{\calU,Y^{\nu:t}}(u,y^{\nu:t})\leq\sum_{j=1}^{|\widetilde{G}|}\max_{u\in\calU}P_{\calU,J}(u,j).
\end{align*}
Hence, for a fixed random variable $U$ that is a randomized function of $I$, we have
\begin{align*}
\frac{\sum_{y^{\nu:t}}\max_{u\in\calU}P_{\calU,Y^{\nu:t}}(u,y^{\nu:t})}{\max_{u\in\calU}P_{\calU}(u)}\leq\frac{\sum_{j=1}^{|\widetilde{G}|}\max_{u\in\calU}P_{\calU,J}(u,j)}{\max_{u\in\calU}P_{\calU}(u)}.
\end{align*}
Using the following equations\cite{issa2016operational},
\begin{align*}
\calL_{\text{max}}(I\to Y^{\nu:t})&=\sup_U \frac{\sum_{y^{\nu:t}}\max_{u\in\calU}P_{\calU,Y^{\nu:t}}(u,y^{\nu:t})}{\max_{u\in\calU}P_{\calU}(u)},\\
\calL_{\text{max}}(I\to J)&=\sup_U \frac{\sum_{j}\max_{u\in\calU}P_{\calU,J}(u,j)}{\max_{u\in\calU}P_{\calU}(u)}.
\end{align*}
to compute maximal leakage, where the supremum is taken over all randomized functions $U$ of $I$, we have 
\begin{align*}
\calL_{\text{max}}(I\to Y^{\nu:t})\leq\calL_{\text{max}}(I\to J).
\end{align*}
The proof is now complete.

\section{Proof of \cref{prop:equiv_conditions}}\label[appendix]{sec:AppProp2}
$(\Rightarrow)$ Suppose the distribution $\phi$ on $\{1,\ldots,n\}$ satisfies \cref{eqn:equiv_prop_1}. Define $\xi:I_1\to \mathbb{R}$ and $\delta:I_1\times I_1 \to\{0,1\}$ such that
\begin{align*}
\xi(i)&=\min_{j\in I_1}\sum_{c=1}^n\phi(c)\KLD{T_cg_i}{T_cg_j}\quad\text{for $j\in I_1$},\\
\delta(j,i)&=\begin{cases}
1\quad\text{if $j=\argmin \sum_{c=1}^n\phi(c)\KLD{T_cg_i}{T_cg_j}$},\\
0\quad\text{otherwise.}\\
\end{cases}
\end{align*}
We can check that $\xi$ and $\delta$ as defined satisfy \cref{eqn:eqn:equiv_prop_2_1,eqn:eqn:equiv_prop_2_2,eqn:eqn:equiv_prop_2_3,eqn:eqn:equiv_prop_2_4}.

$(\Leftarrow)$Now suppose that there exist functions $\xi:I_1\to \mathbb{R}$ and $\delta:I_1\times I_1 \to\{0,1\}$ such that the distribution $\phi$ on $\{1,\ldots,n\}$ satisfies \cref{eqn:eqn:equiv_prop_2_1,eqn:eqn:equiv_prop_2_2,eqn:eqn:equiv_prop_2_3,eqn:eqn:equiv_prop_2_4}. Fix $i\in I_1$. From \eqref{eqn:eqn:equiv_prop_2_3}, we have 
\begin{align*}
\sum_{c=1}^n\phi(c)\KLD{T_cg_i}{T_cg_j}\geq \xi(i),
\end{align*} for each $j\in I_1$. Taking minimum over $j\in I_1$, we obtain
\begin{align*}
\min_{j\in I_1}\sum_{c=1}^n\phi(c)\KLD{T_cg_i}{T_cg_j}\geq \xi(i).
\end{align*} From \eqref{eqn:eqn:equiv_prop_2_2}, we have 
\begin{align*}
\sum_{j\in I_1}\delta(j,i)=1.
\end{align*} Since $\delta(j,i)\in\{0,1\}$, there exists a unique $j'$ such that $\delta(j',i)=1$. Putting this together with \cref{eqn:eqn:equiv_prop_2_4}, we obtain \begin{align*}
\min_{j\in I_1}\sum_{c=1}^n\phi(c)\KLD{T_cg_i}{T_cg_j}=\xi(i).
\end{align*}
By \cref{eqn:eqn:equiv_prop_2_1}, we obtain\begin{align*}
\max_{i\in I_1} \min_{j\in I_1}\sum_{c=1}^n\phi(c)\KLD{T_cg_i}{T_cg_j}\leq \epsilon_1.
\end{align*}
The proposition is proved.

\section{Proof of \cref{prop:optimality}}\label[appendix]{sec:AppProp3}
Let $q_1^\dagger,\ldots,q_K^\dagger$ be an optimal solution to Problem~\eqref{eqn:optimize_relaxed_formulation} under the decentralized QCD setting, $P_{J|I}^\dagger$ be the corresponding conditional probability distribution of $J$ given $I$, and $\mu^\dagger$ be the corresponding optimal value. 

Since the observations obtained by the sensors are mutually independent,
we have 
\begin{align*}
\E[\KLD{\widetilde{g}_I}{\widetilde{f}}]=\sum_{k=1}^K\E[\KLD{T_{q_k}g_{k,I}}{T_{q_k}f_{k}}].
\end{align*}
By our construction of $q^*_k(P_{J|I}^\dagger)$,  we have 
\begin{align}
\label{eqn:prop2_1_1}
\begin{split}
&\E[\KLD{T_{q_k^*\left(P_{J|I}^\dagger\right)} g_{k,I}}{T_{q_k^*\left(P_{J|I}^\dagger\right)} f_k}]
\\
&\quad\geq \E[\KLD{T_{q_k^\dagger} g_{k,I}}{T_{q_k^\dagger} f_k}],
\end{split}
\end{align}
for each $k\in\{1,\ldots,K\}$. Let $\mu^*$ be the value achieved by the cost function in Problem~\eqref{eqn:optimize_relaxed_formulation} corresponding to $\{q_1^*(P_{J|I}^*),\ldots,q_k^*(P_{J|I}^*)\}$ and we obtain
\begin{align}
&\mu^*=\sum_{k=1}^K\E[\KLD{T_{q_k^*\left(P_{J|I}^*\right)} g_{k,I}}{T_{q_k^*\left(P_{J|I}^*\right)} f_k}]\label{eqn:prop2_1}\\
&\quad\geq \sum_{k=1}^K\E[\KLD{T_{q_k^*\left(P_{J|I}^\dagger\right)} g_{k,I}}{T_{q_k^*\left(P_{J|I}^\dagger\right)} f_k}]\label{eqn:prop2_2}\\
&\quad\geq\sum_{k=1}^K \E[\KLD{T_{q_k^\dagger} g_{k,I}}{T_{q_k^\dagger} f_k}]=\mu^\dagger\label{eqn:prop2_3}
\end{align}
where the inequality from \eqref{eqn:prop2_1} to \eqref{eqn:prop2_2} is a consequence of our choice of $P_{J|I}^*$ in \eqref{eqn:cases_2} and the inequality from \eqref{eqn:prop2_2} to \eqref{eqn:prop2_3} is obtained by summing \eqref{eqn:prop2_1_1} over $k\in\{1,\ldots,K\}$. Since $P_{J|I}^*$ also satisfies \begin{align*}\calL(I\to J)\leq\epsilon,\end{align*} $\{q_k^*(P_{J|I}^*),\ldots,q_k^*(P_{J|I}^*)\}$ is also an optimal solution to Problem~\eqref{eqn:optimize_relaxed_formulation}. The proof is now complete.	

\bibliographystyle{IEEETran}
\bibliography{IEEEabrv,StringDefinitions,refs}

\end{document}

%% file: preamble.tex
\usepackage[T1]{fontenc}
\usepackage{amsmath,amssymb,amsfonts,mathrsfs,bm}
\usepackage{mathtools}
\usepackage{amsthm}
\usepackage{cite}
\usepackage{array}
\usepackage[shortlabels]{enumitem}
\usepackage{graphicx}
\usepackage{epstopdf}
\usepackage{url}
\usepackage{color}
\usepackage{multirow}
\usepackage[table]{xcolor}
\usepackage[normalem]{ulem}
\usepackage{array}
\newcolumntype{L}[1]{>{\raggedright\let\newline\\\arraybackslash\hspace{0pt}}m{#1}}
\newcolumntype{C}[1]{>{\centering\let\newline\\\arraybackslash\hspace{0pt}}m{#1}}
\newcolumntype{R}[1]{>{\raggedleft\let\newline\\\arraybackslash\hspace{0pt}}m{#1}}
\usepackage{subcaption}
\usepackage{xparse}



\usepackage{glossaries}
\newacronym{wrt}{w.r.t.}{with respect to}
\newacronym{RHS}{R.H.S.}{right-hand side}
\newacronym{LHS}{L.H.S.}{left-hand side}
\newacronym{iid}{i.i.d.}{independent and identically distributed}
\newacronym{MIMO}{MIMO}{mulitple-input multiple-output}
\newacronym{AOA}{AOA}{angle-of-arrival}
\newacronym{AOD}{AOD}{angle-of-departure}
\newacronym{LOS}{LOS}{line-of-sight}
\newacronym{NLOS}{NLOS}{non-line-of-sight}
\newacronym{TOA}{TOA}{time-of-arrival}
\newacronym{TDOA}{TDOA}{time-difference-of-arrival}
\newacronym{RSS}{RSS}{received signal strength}
\newacronym{GNSS}{GNSS}{Global Navigation Satellite System}

\usepackage{float}

\ifx\notloadhyperref\undefined
	\ifx\loadbibentry\undefined
		\usepackage[hidelinks,hypertexnames=false]{hyperref} 
	\else
		\usepackage{bibentry}
		\makeatletter\let\saved@bibitem\@bibitem\makeatother
		\usepackage[hidelinks,hypertexnames=false]{hyperref}
		\makeatletter\let\@bibitem\saved@bibitem\makeatother
	\fi
\else
	\relax
\fi

\usepackage[capitalize]{cleveref}
\crefname{equation}{}{}
\Crefname{equation}{}{}
\crefname{claim}{claim}{claims}
\crefname{step}{step}{steps}
\crefname{line}{line}{lines}
\crefname{dmath}{}{}
\crefname{dseries}{}{}
\crefname{dgroup}{}{}
\crefname{Theorem}{Theorem}{Theorems}
\crefname{Corollary}{Corollary}{Corollaries}
\crefname{Proposition}{Proposition}{Propositions}
\crefname{Lemma}{Lemma}{Lemmas}
\crefname{Definition}{Definition}{Definitions}
\crefname{Example}{Example}{Examples}
\crefname{Assumption}{Assumption}{Assumptions}
\crefname{Remark}{Remark}{Remarks}
\crefname{Rem}{Remark}{Remarks}
\crefname{remarks}{Remarks}{Remarks}
\crefname{Theorem_A}{Theorem}{Theorems}
\crefname{Corollary_A}{Corollary}{Corollaries}
\crefname{Proposition_A}{Proposition}{Propositions}
\crefname{Lemma_A}{Lemma}{Lemmas}
\crefname{Definition_A}{Definition}{Definitions}

\usepackage{algorithm,algorithmic}

\ifx\loadbreqn\undefined
	\relax
\else
	\usepackage{breqn} 
\fi

\ifx\renewtheorem\undefined
\ifx\useTheoremCounter\undefined
\newtheorem{Theorem}{Theorem}
\newtheorem{Corollary}{Corollary}
\newtheorem{Proposition}{Proposition}

\else

\newtheorem{Proposition}[theorem]{Proposition}
\fi

\fi

\theoremstyle{remark}

\theoremstyle{plain}

\newcommand{\calA}{\mathcal{A}}
\newcommand{\calB}{\mathcal{B}}
\newcommand{\calC}{\mathcal{C}}

\newcommand{\calK}{\mathcal{K}}
\newcommand{\calL}{\mathcal{L}}

\newcommand{\calQ}{\mathcal{Q}}

\newcommand{\calU}{\mathcal{U}}

\newcommand{\calX}{\mathcal{X}}
\newcommand{\calY}{\mathcal{Y}}

\newcommand{\bX}{\mathbf{X}}

\newcommand{\bbN}{\mathbb{N}}

\DeclareSymbolFont{bsfletters}{OT1}{cmss}{bx}{n}
\DeclareSymbolFont{ssfletters}{OT1}{cmss}{m}{n}
\DeclareMathSymbol{\bsfGamma}{0}{bsfletters}{'000}
\DeclareMathSymbol{\ssfGamma}{0}{ssfletters}{'000}
\DeclareMathSymbol{\bsfDelta}{0}{bsfletters}{'001}
\DeclareMathSymbol{\ssfDelta}{0}{ssfletters}{'001}
\DeclareMathSymbol{\bsfTheta}{0}{bsfletters}{'002}
\DeclareMathSymbol{\ssfTheta}{0}{ssfletters}{'002}
\DeclareMathSymbol{\bsfLambda}{0}{bsfletters}{'003}
\DeclareMathSymbol{\ssfLambda}{0}{ssfletters}{'003}
\DeclareMathSymbol{\bsfXi}{0}{bsfletters}{'004}
\DeclareMathSymbol{\ssfXi}{0}{ssfletters}{'004}
\DeclareMathSymbol{\bsfPi}{0}{bsfletters}{'005}
\DeclareMathSymbol{\ssfPi}{0}{ssfletters}{'005}
\DeclareMathSymbol{\bsfSigma}{0}{bsfletters}{'006}
\DeclareMathSymbol{\ssfSigma}{0}{ssfletters}{'006}
\DeclareMathSymbol{\bsfUpsilon}{0}{bsfletters}{'007}
\DeclareMathSymbol{\ssfUpsilon}{0}{ssfletters}{'007}
\DeclareMathSymbol{\bsfPhi}{0}{bsfletters}{'010}
\DeclareMathSymbol{\ssfPhi}{0}{ssfletters}{'010}
\DeclareMathSymbol{\bsfPsi}{0}{bsfletters}{'011}
\DeclareMathSymbol{\ssfPsi}{0}{ssfletters}{'011}
\DeclareMathSymbol{\bsfOmega}{0}{bsfletters}{'012}
\DeclareMathSymbol{\ssfOmega}{0}{ssfletters}{'012}

\DeclareMathOperator*{\argmax}{arg\,max}
\DeclareMathOperator*{\argmin}{arg\,min}

\DeclareMathOperator*{\esssup}{ess\,sup}

\newcommand{\qednew}{\nobreak \ifvmode \relax \else
      \ifdim\lastskip<1.5em \hskip-\lastskip
      \hskip1.5em plus0em minus0.5em \fi \nobreak
      \vrule height0.75em width0.5em depth0.25em\fi}

\newcommand{\nn}{\nonumber\\}

\newcommand{\KLD}[2]{{D({#1}\ ||\ {#2})}}

\newcommand{\cond}[2]{\left. {#1}\, \middle| \, {#2} \right.}

\DeclareDocumentCommand \P { g d() g } {%
	\IfNoValueTF {#3} 
	{%
		\IfNoValueTF {#1} 
		{%
			\IfNoValueTF {#2}
			{%
				\mathbb{P}%
			}%
			{%
				\mathbb{P}\left({#2}\right)%
			}%
		}%
		{%
			\IfNoValueTF {#2}
			{%
				\mathbb{P}_{#1}%
			}%
			{%
				\mathbb{P}_{#1}\left({#2}\right)%
			}%
		}%
	}%
	{%
		\IfNoValueTF {#1} 
		{%
			\mathbb{P}\left(\cond{#2}{#3}\right)%
		}%
		{%
			\mathbb{P}_{#1}\left(\cond{#2}{#3}\right)%
		}%
	}%
}

\DeclareDocumentCommand \E { g o g } {%
	\IfNoValueTF {#3} 
	{%
		\IfNoValueTF {#1} 
		{%
			\IfNoValueTF {#2}
			{%
				\mathbb{E}%
			}%
			{%
				\mathbb{E}\left[{#2}\right]%
			}%
		}%
		{%
			\IfNoValueTF {#2}
			{%
				\mathbb{E}_{#1}%
			}%
			{%
				\mathbb{E}_{#1}\left[{#2}\right]%
			}%
		}%
	}%
	{%
		\IfNoValueTF {#1} 
		{%
			\mathbb{E}\left[\cond{#2}{#3}\right]%
		}%
		{%
			\mathbb{E}_{#1}\left[\cond{#2}{#3}\right]%
		}%
	}%
}

\definecolor{gray90}{gray}{0.9}

\newcommand{\msout}[1]{\text{\color{green} \sout{\ensuremath{#1}}}}
\newcommand{\del}[1]{{\color{green}\ifmmode \msout{#1}\else\sout{#1}\fi}}

\newcommand{\hide}[1]{}

\renewcommand{\figurename}{Fig.}
\newcommand{\figref}[1]{\figurename~\ref{#1}}
\graphicspath{{./Figures/}} 